\documentclass[12pt]{iopart}

\usepackage{amsmath}
\usepackage{amssymb}
\usepackage[english]{babel}
\usepackage[pdfborder={0 0 0},breaklinks=true]{hyperref}
\usepackage{cleveref} % hyperref must be loaded first
\usepackage{microtype}
\usepackage{mathtools}
\PassOptionsToPackage{dvipsnames}{xcolor}
\usepackage{tikz}
\usepackage{xspace}

\newcommand{\tcolorboxpath}{texmf/tcolorbox}
\usepackage[skins]{\tcolorboxpath/tcolorbox}

\usetikzlibrary{calc}

\usepackage{texmf/formulautils}

\crefname{section}{sec.}{sections}
\Crefname{equation}{Eq.}{Eqns.}

\NewDocumentCommand{\name}{m}{%
	\IfEqCase{#1}{%
		{Poincare}{Poincar\'{e}}%
		{Ampere}{Amp\`{e}re}%
	}%
	[\PackageError{}{Undefined name:}{}]%
}

\NewDocumentCommand{\abbr}{m}{%
	\IfEqCase{#1}{%
		{abbrev}{abbrev.\xspace}%
		{AD}{A.\,D.\xspace}%
		{BC}{B.\,C.\xspace}%
		{cf}{cf.\xspace}%
		{cmp}{cmp.\xspace}%
		{dof}{d.\,o.\,f.\xspace}%
		{eg}{e.\,g.\xspace}%
		{et_al}{et\,al.\xspace}%
		{etc}{etc.\xspace}%
		{fig}{fig.\xspace}%
		{ie}{i.\,e.\xspace}%
		{p}{p.\xspace}%
		{sec}{sec.\xspace}%
		{vs}{vs.\xspace}%
		{wlog}{w.\,l.\,o.\,g.\xspace}%
		{wrt}{w.\,r.\,t.\xspace}%
	}%
	[\PackageError{}{Undefined abbreviation:}{}]%
}

\NewDocumentCommand{\Abbr}{m}{%
	\IfEqCase{#1}{%
		{abbrev}{Abbrev.\xspace}%
		{AD}{A.\,D.\xspace}%
		{BC}{B.\,C.\xspace}%
		{cf}{Cf.\xspace}%
		{cmp}{Cmp.\xspace}%
		{dof}{D.\,o.\,f.\xspace}%
		{eg}{E.\,g.\xspace}%
		{fig}{Fig.\xspace}%
		{ie}{I.\,e.\xspace}%
		{p}{P.\xspace}%
		{sec}{Sec.\xspace}%
		{vs}{Vs.\xspace}%
		{wlog}{W.\,l.\,o.\,g.\xspace}%
		{wrt}{W.\,r.\,t.\xspace}%
	}%
	[\PackageError{}{Undefined abbreviation:}{}]%
}

\NewDocumentCommand{\mconst}{m}{%
	\IfEqCase{#1}{%
		{e}{\mathrm{e}}%
		{i}{\mathrm{i}}%
		{pi}{\pi}%
		{sigma1}{\hat{\sigma}_1}% 1st Pauli spin matrix
		{sigma2}{\hat{\sigma}_2}% 2nd Pauli spin matrix
		{sigma3}{\hat{\sigma}_3}% 3rd Pauli spin matrix
		{vec_sigma}{\hat{\vec{\sigma}}}% vector of the Pauli spin matrices
	}%
	[\PackageError{}{Undefined mathematical constant:}{}]%
}

\NewDocumentCommand{\pconst}{m}{%
	\IfEqCase{#1}{%
		{c}{\mathrm{c}}% speed of light in vacuum
		{e}{\mathrm{e}}% elementary charge
		{eps0}{\epsilon_0}% vacuum permittivity
		{h}{\mathrm{h}}% Planck constant
		{hbar}{\hbar}% reduced Planck constant
		{kB}{\mathrm{k_B}}% Boltzmann constant
		{me}{m_\mathrm{e}}% electron mass
		{mu0}{\mu_0}% magnetic field constant (name correct?)
	}%
	[\PackageError{}{Undefined physical constant:}{}]%
}

\NewDocumentCommand{\pquant}{m}{%
	\IfEqCase{#1}{%
		{E_F}{E_\mathrm{F}}% Fermi energy
		{H_bsf}{\hat{H}_\mathrm{bsf}}%
		{H_fsf}{\hat{H}_\mathrm{fsf}}%
		{H_magn}{\hat{H}_\mathrm{magn}}%
		{H_std}{\hat{H}_\mathrm{std}}%
		{H_Zeeman}{\hat{H}_\mathrm{Zeeman}}% Zeeman splitting Hamiltonian
		{Phi_L}{\Phi_\mathrm{L}}% gauge condition in the left-handed regions
		{Phi_R}{\Phi_\mathrm{R}}% gauge condition in the right-handed regions
		{psi_L}{\Psi_\mathrm{L}}% the left-circular component
		{psi_R}{\Psi_\mathrm{R}}% the right-circular component
		{polar_azimuth}{\theta}% polarization azimuth (1/2 times the azimuth angle in the Poincare sphere)
		{polar_altitude}{\chi}% polarization altitude (1/2 times the altitude angle in the Poincare sphere)
		{r_f}{\vec{r}_\mathrm{f}}% final position
		{r_i}{\vec{r}_\mathrm{i}}% initial position
		{S_T}{\hat{S}_{\oper{T}}}% unitary transformation of the (physical) time-reversal operator
		{S_Th}{\hat{S}_{\oper{Th}}}% unitary transformation of the (abstract) time-reversal operator
		{S_PH}{\hat{S}_{\oper{PH}}}% unitary transformation of the particle-hole operator
		{S_Chir}{\hat{S}_{\oper{Chir}}}% unitary transformation which represents the chiral operator
		{Sx}{S_x}% x component in the Bloch sphere
		{Sy}{S_y}% y component in the Bloch sphere
		{Sz}{S_z}% z component in the Bloch sphere
		{S0}{S_0}% 0th Stokes parameter (component in the Poincare sphere)
		{S1}{S_1}% 1st Stokes parameter (component in the Poincare sphere)
		{S2}{S_2}% 2nd Stokes parameter (component in the Poincare sphere)
		{S3}{S_3}% 3rd Stokes parameter (component in the Poincare sphere)
		{t_f}{t_\mathrm{f}}% final time
		{t_i}{t_\mathrm{i}}% initial time
		{T_ferm}{\mathcal{T}_\mathrm{ferm}}% fermionic time-reversal operator
		{T_bos}{\mathcal{T}_\mathrm{bos}}% bosonic time-reversal operator
		{T_bipartite}{\tau}% fermionic-like time-reversal operator with a relative minus sign
		{t^bsf}{t^\mathrm{(bsf)}}% bosonic spin-flip hopping
		{t^fsf}{t^\mathrm{(fsf)}}% fermionic spin-flip hopping
		{t^magn}{t^\mathrm{(magn)}}% "magnetic" hopping
		{t^std}{t^\mathrm{(std)}}% standard SOC hopping
		{t^dd}{t^{(\downarrow\downarrow)}}% hopping down->down
		{t^du}{t^{(\downarrow\uparrow)}}% hopping down->up
		{t^ud}{t^{(\uparrow\downarrow)}}% hopping up->down
		{t^uu}{t^{(\uparrow\uparrow)}}% hopping up->up
		{t^xx}{t^{(xx)}}% hopping x->x
		{t^xy}{t^{(xy)}}% hopping x->y
		{t^yx}{t^{(yx)}}% hopping y->x
		{t^yy}{t^{(yy)}}% hopping y->y
		{W_GSOC}{W_\mathrm{GSOC}}% GSOC coupling constant
		{W_iso}{W_\mathrm{iso}}% spin-isotropic coupling constant
		{W_lon}{W_\mathrm{lon}}% longitudinal coupling constant
		{W_tra}{W_\mathrm{tra}}% transversal coupling constant
		{W_Zeeman}{W_\mathrm{Zeeman}}% Zeeman splitting coupling constant
	}%
	[\PackageError{}{Undefined physical quantity:}{}]%
}

\NewDocumentCommand{\mathset}{m}{%
	\IfEqCase{#1}{%
		{C}{\mathbb{C}}
		{C2}{\mathbb{C}^2}
		{H}{\mathcal{H}} % Hilbert space
		{N}{\mathbb{N}}
		{O1}{\mathsf{O}(1)}
		{O2}{\mathsf{O}(2)}
		{O3}{\mathsf{O}(3)}
		{Q}{\mathbb{Q}}
		{R}{\mathbb{R}}
		{R+}{\mathbb{R}^+}
		{R0+}{\mathbb{R}_0^+}
		{R2}{\mathbb{R}^2}
		{R3}{\mathbb{R}^3}
		{R4}{\mathbb{R}^4}
		{SO1}{\mathsf{SO}(1)}
		{SO2}{\mathsf{SO}(2)}
		{SO3}{\mathsf{SO}(3)}
		{SU1}{\mathsf{SU}(1)}
		{SU2}{\mathsf{SU}(2)}
		{SU3}{\mathsf{SU}(3)}
		{U1}{\mathsf{U}(1)}
		{U2}{\mathsf{U}(2)}
		{U3}{\mathsf{U}(3)}
		{Z}{\mathbb{Z}}
		{Z2}{\mathbb{Z}^2}
		{Z_2}{\mathbb{Z}_2}
	}%
	[\PackageError{}{Undefined name:}{}]%
}

\NewDocumentCommand{\opfun}{mm}{%
	\IfEqCase{#2}{%
		{cc}{\conj{#1}} % complex conjugation
		{hc}{{#1}^\dagger} % Hermitian conjugation
		{inv}{{#1}^{-1}} % inverse
		{transposed}{{#1}^\mathsf{T}} % transposed
	}%
	[\PackageError{}{Undefined name:}{}]%
}

\NewDocumentCommand{\term}{m}{%
	\IfEqCase{#1}{%
		{cc}{\mathsf{c.\,c.}} % complex conjugated
		{const}{\mathsf{const}} % constant
		{hc}{\mathsf{h.\,c.}} % Hermitian conjugated
	}%
	[\PackageError{}{Undefined name:}{}]%
}

\NewDocumentCommand{\neighbors}{mm}{%
	\IfEqCase{#1}{%
		{1}{\langle#2\rangle}%
		{2}{\llangle#2\rrangle}%
	}%
	[\PackageError{}{Undefined neighbor level:}{}]%
}

\definecolor{mycyan}{rgb}{0,1,1}

\NewDocumentCommand{\multiarrow}{O{draw=black}mmmmmmm}{%
	\path[#1] ({(#2)-0.5*(#4)},{(#3)+0.5*(#6)})
			\foreach \counter in {1,...,#8} {to ({(#2)+(\counter/((#8)+1)*(1+(#5))-(#5)-0.5)*(#4)},{(#3)+0.5*(#6)}) to ({(#2)+(\counter/((#8)+1)*(1+(#5))-(#5)-0.5)*(#4)},{(#3)+0.5*(#7)}) to ({(#2)+(\counter/((#8)+1)*(1+(#5))-0.5)*(#4)},{(#3)+0.5*(#6)})}
			to ({(#2)+0.5*(#4)},{(#3)+0.5*(#6)}) to ({(#2)+0.5*(#4)},{(#3)-0.5*(#6)})
			\foreach \counter in {#8,...,1} {to ({(#2)+(\counter/((#8)+1)*(1+(#5))-0.5)*(#4)},{(#3)-0.5*(#6)}) to ({(#2)+(\counter/((#8)+1)*(1+(#5))-(#5)-0.5)*(#4)},{(#3)-0.5*(#7)}) to ({(#2)+(\counter/((#8)+1)*(1+(#5))-(#5)-0.5)*(#4)},{(#3)-0.5*(#6)})}
			to ({(#2)-0.5*(#4)},{(#3)-0.5*(#6)}) -- cycle;
}

\NewDocumentCommand{\winding}{O{}mmmmmm}{%
	\foreach \tick in {0,...,#6}{
		\draw[#1] ($({#2+(\tick/#6-0.5)*#4},#3)-(#7*\tick/#6*180+90:0.5*#5)$) to ($({#2+(\tick/#6-0.5)*#4},#3)+(#7*\tick/#6*180+90:0.5*#5)$);
	}
}

\definecolor{bps_sphere_background}{gray}{0.85}
\definecolor{bps_sphere_contours}{gray}{0.35}
\definecolor{bps_angles}{rgb}{0.54,0.1,0.65}
\definecolor{bps_north_pole}{rgb}{1,0.25,0}
\definecolor{bps_equator}{rgb}{0,0.5,0}
\definecolor{bps_south_pole}{rgb}{0,0,1}

\newcommand{\BPSPROJ}[3]{{(#1)*cos(\azimut)-(#2)*sin(\azimut)},{(#1)*sin(\elevation)*sin(\azimut)+(#2)*sin(\elevation)*cos(\azimut)+(#3)*cos(\elevation)}}
\newcommand{\BPSCOND}[3]{(#1)*(#2)+(1-(#1))*(#3)}

\NewDocumentCommand{\bpsDrawSphere}{O{meridian}O{equator}O{margin}m}{
	\IfEqCase{#4}{%
		{opaque}{
			\fill[bps_sphere_background,ultra thick] (0,0) circle (\sphereradius);
		}%
		{shaded}{%
			\pgfmathsetmacro{\SHADEDSPHEREIMAGESCALE}{0.1108*\sphereradius}
			\node at (0,0) {\includegraphics[scale=\SHADEDSPHEREIMAGESCALE]{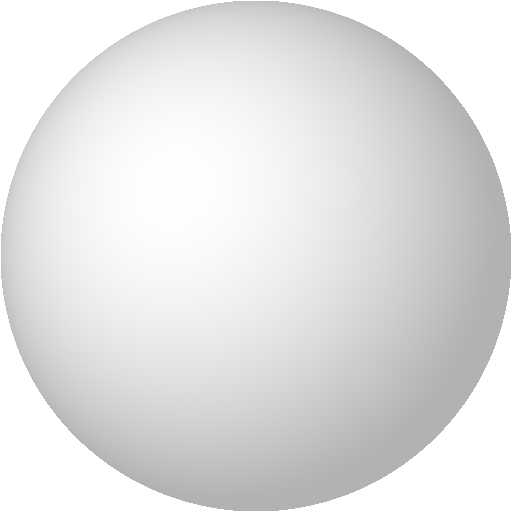}};
		}%
		{transparent}{}%
	}[\PackageError{}{Undefined option (expected "opaque, "shaded" or "transparent"):}{}]%
	\IfEqCase{#3}{%
		{margin}{\draw[bps_sphere_contours,ultra thick] (0,0) circle (\sphereradius);}%
		{nomargin}{}%
	}[\PackageError{}{Undefined option (expected "margin" or "nomargin"):}{}]%
	\IfEqCase{#2}{%
		{equator}{%
			\draw[bps_sphere_contours,rotate={\BPSCOND{cos(\elevation)>0}{180}{0}}] (\sphereradius,0) arc (0:180:{\sphereradius} and {(\sphereradius)*sin(\elevation)});%
			\draw[bps_sphere_contours,rotate={\BPSCOND{cos(\elevation)>0}{0}{180}},dashed] (\sphereradius,0) arc (0:180:{\sphereradius} and {(\sphereradius)*sin(\elevation)});%
		}%
		{noequator}{}%
	}[\PackageError{}{Undefined option (expected "equator" or "noequator"):}{}]%
	\IfEqCase{#1}{%
		{meridian}{%
			\draw[bps_sphere_contours,rotate={\BPSCOND{cos(\azimut)^2-sin(\elevation)^2*sin(\azimut)^2-cos(\elevation)^2>0}{0}{90}+\BPSCOND{sin(2*(\elevation))*sin(2*(\azimut))>0}{0}{180}+0.5*atan(sin(\elevation)*sin(2*(\azimut))/(cos(\azimut)^2-sin(\elevation)^2*sin(\azimut)^2-cos(\elevation)^2))}] (\sphereradius,0) arc (0:180:{\sphereradius} and {(\sphereradius)*sqrt(cos(\azimut)^2+sin(\elevation)^2*sin(\azimut)^2+cos(\elevation)^2-1)});%
			\draw[bps_sphere_contours,rotate={\BPSCOND{cos(\azimut)^2-sin(\elevation)^2*sin(\azimut)^2-cos(\elevation)^2>0}{0}{90}+\BPSCOND{sin(2*(\elevation))*sin(2*(\azimut))>0}{180}{0}+0.5*atan(sin(\elevation)*sin(2*(\azimut))/(cos(\azimut)^2-sin(\elevation)^2*sin(\azimut)^2-cos(\elevation)^2))},dashed] (\sphereradius,0) arc (0:180:{\sphereradius} and {(\sphereradius)*sqrt(cos(\azimut)^2+sin(\elevation)^2*sin(\azimut)^2+cos(\elevation)^2-1)});%
			\draw[bps_sphere_contours,rotate={\BPSCOND{sin(\azimut)^2-sin(\elevation)^2*cos(\azimut)^2-cos(\elevation)^2>0}{0}{90}+\BPSCOND{sin(2*(\elevation))*sin(2*(\azimut))>0}{180}{0}-0.5*atan(sin(\elevation)*sin(2*(\azimut))/(sin(\azimut)^2-sin(\elevation)^2*cos(\azimut)^2-cos(\elevation)^2))}] (\sphereradius,0) arc (0:180:{\sphereradius} and {(\sphereradius)*sqrt(sin(\azimut)^2+sin(\elevation)^2*cos(\azimut)^2+cos(\elevation)^2-1)});%
			\draw[bps_sphere_contours,rotate={\BPSCOND{sin(\azimut)^2-sin(\elevation)^2*cos(\azimut)^2-cos(\elevation)^2>0}{0}{90}+\BPSCOND{sin(2*(\elevation))*sin(2*(\azimut))>0}{0}{180}-0.5*atan(sin(\elevation)*sin(2*(\azimut))/(sin(\azimut)^2-sin(\elevation)^2*cos(\azimut)^2-cos(\elevation)^2))},dashed] (\sphereradius,0) arc (0:180:{\sphereradius} and {(\sphereradius)*sqrt(sin(\azimut)^2+sin(\elevation)^2*cos(\azimut)^2+cos(\elevation)^2-1)});%
		}%
		{nomeridian}{}%
	}[\PackageError{}{Undefined option (expected "meridian" or "nomeridian"):}{}]%
}

\NewDocumentCommand{\bpsDrawAxes}{oO{0.0}m}{
	\draw[thick,-latex] (0,0) to (\BPSPROJ{(#3)*(\sphereradius)}{0}{0});
	\draw[thick,-latex] (0,0) to (\BPSPROJ{0}{(#3)*(\sphereradius)}{0});
	\draw[thick,-latex] (0,0) to (\BPSPROJ{0}{0}{(#3)*(\sphereradius)});
	\IfNoValueF{#1}{%
		\IfEqCase{#1}{%
			{bloch}{%
				\node at (\BPSPROJ{(#3)*(\sphereradius)+(#2)}{0}{0}) {$\pquant{Sx}$};
				\node at (\BPSPROJ{0}{(#3)*(\sphereradius)+(#2)}{0}) {$\pquant{Sy}$};
				\node at (\BPSPROJ{0}{0}{(#3)*(\sphereradius)+(#2)}) {$\pquant{Sz}$};
			}%
			{poincare}{%
				\node at (\BPSPROJ{(#3)*(\sphereradius)+(#2)}{0}{0}) {$\pquant{S1}$};
				\node at (\BPSPROJ{0}{(#3)*(\sphereradius)+(#2)}{0}) {$\pquant{S2}$};
				\node at (\BPSPROJ{0}{0}{(#3)*(\sphereradius)+(#2)}) {$\pquant{S3}$};
			}%
		}[\PackageError{}{Undefined option (expected "bloch" or "poincare"):}{}]%
	}%
}

\NewDocumentCommand{\bpsDrawAngles}{oO{0}O{0.5}moO{0}O{0.5}mm}{
	\pgfmathsetmacro{\SPHEREAZIMUTANGLEROTATION}{\BPSCOND{cos(\azimut)^2-sin(\elevation)^2*sin((\azimut)+2*(#4))^2-cos(\elevation)^2>0}{0}{90}+\BPSCOND{sin(2*(\elevation))*sin(2*(\azimut)+4*(#4))>0}{0}{180}+0.5*atan(sin(\elevation)*sin(2*(\azimut)+4*(#4))/(cos((\azimut)+2*(#4))^2-sin(\elevation)^2*sin((\azimut)+2*(#4))^2-cos(\elevation)^2))}
	\pgfmathsetmacro{\SPHEREAZIMUTANGLESTART}{atan(tan(atan(tan((\azimut)+2*(#4))*sin(\elevation))-(\SPHEREAZIMUTANGLEROTATION))/sqrt(cos((\azimut)+2*(#4))^2+sin(\elevation)^2*sin((\azimut)+2*(#4))^2+cos(\elevation)^2-1))};
	\pgfmathsetmacro{\SPHEREAZIMUTANGLEFINISH}{180+atan(tan(atan(tan((\azimut)+2*(#4))*sin(\elevation)+cos(\elevation)*tan(2*(#8))/cos((\azimut)+2*(#4)))-(\SPHEREAZIMUTANGLEROTATION))/sqrt(cos((\azimut)+2*(#4))^2+sin(\elevation)^2*sin((\azimut)+2*(#4))^2+cos(\elevation)^2-1))};
	\draw[bps_angles,dotted] (0,0) to (\BPSPROJ{(\sphereradius)*cos(2*(#4))}{(\sphereradius)*sin(2*(#4))}{0});
	\draw[bps_angles] (0,0) to (\BPSPROJ{(\sphereradius)*cos(2*(#4))*cos(2*(#8))}{(\sphereradius)*sin(2*(#4))*cos(2*(#8))}{(\sphereradius)*sin(2*(#8))});
	\fill[bps_angles] (\BPSPROJ{(\sphereradius)*cos(2*(#4))*cos(2*(#8))}{(\sphereradius)*sin(2*(#4))*cos(2*(#8))}{(\sphereradius)*sin(2*(#8))}) circle (0.1);
	\draw[bps_angles,-latex] (\BPSPROJ{(#9)*(\sphereradius)}{0}{0}) arc (\azimut:(\azimut)+2*(#4):{(#9)*(\sphereradius)} and {(#9)*(\sphereradius)*sin(\elevation)});
	\draw[bps_angles,-latex,rotate=\SPHEREAZIMUTANGLEROTATION] ({(#9)*(\sphereradius)*cos(\SPHEREAZIMUTANGLESTART)},{(#9)*(\sphereradius)*sin(\SPHEREAZIMUTANGLESTART)*sqrt(cos((\azimut)+2*(#4))^2+sin(\elevation)^2*sin((\azimut)+2*(#4))^2+cos(\elevation)^2-1))}) arc (\SPHEREAZIMUTANGLESTART:\SPHEREAZIMUTANGLEFINISH:{(#9)*(\sphereradius)} and {(#9)*(\sphereradius)*sqrt(cos((\azimut)+2*(#4))^2+sin(\elevation)^2*sin((\azimut)+2*(#4))^2+cos(\elevation)^2-1)});
	\IfNoValueF{#1}{\node[bps_angles] at (\BPSPROJ{((#9)*(\sphereradius)+(#2))*cos(2*(#3)*(#4))}{((#9)*(\sphereradius)+(#2))*sin(2*(#3)*(#4))}{0}) {#1};}
	\IfNoValueF{#5}{\node[bps_angles] at (\BPSPROJ{((#9)*(\sphereradius)+(#6))*cos(2*(#4))*cos(2*(#7)*(#8))}{((#9)*(\sphereradius)+(#6))*sin(2*(#4))*cos(2*(#7)*(#8))}{((#9)*(\sphereradius)+(#6))*sin(2*(#7)*(#8))}) {#5};}
	\draw[bps_angles,|-|,shift={({atan2(sin(\azimut+2*#4)*sin(2*#8)*sin(\elevation)+sin(2*#8)*cos(\elevation), cos(\azimut+2*#4)*cos(2*#8))+90}:0.3)}] (0,0) to node[above] {$S_0$} (\BPSPROJ{(\sphereradius)*cos(2*(#4))*cos(2*(#8))}{(\sphereradius)*sin(2*(#4))*cos(2*(#8))}{(\sphereradius)*sin(2*(#8))});
}

\NewDocumentCommand{\bpsDrawPoincareDeco}{O{0.75}mmm}{
	\path[fill=white,draw=black] (\BPSPROJ{0}{0}{\sphereradius}) circle (#2);
	\path[fill=white,draw=black] (\BPSPROJ{0}{0}{-(\sphereradius)}) circle (#2);
	\foreach \angle in {0,...,3}{
		\path[fill=white,draw=black] (\BPSPROJ{(\sphereradius)*cos(90*\angle)}{(\sphereradius)*sin(90*\angle)}{0}) circle (#2);
	}
	
	\pgfmathsetmacro{\LINEARMOTIONLENGTH}{(#1)*(#2)}
	\pgfmathsetmacro{\CIRCULARMOTIONRADIUS}{sqrt(0.5)*(#1)*(#2)}
	\pgfmathsetmacro{\ARROWHALFWIDTH}{0.5*(#3)}
	\pgfmathsetmacro{\ARROWHALFLENGTH}{0.5*(#4)}
	
	\draw[bps_north_pole] (\BPSPROJ{0}{0}{\sphereradius}) circle ({(#1)*(#2)*sqrt(0.5)});
	\foreach \angle in {0,...,3}{
		\fill[bps_north_pole] ($(\BPSPROJ{0}{0}{\sphereradius})+(90*\angle+45:\CIRCULARMOTIONRADIUS)+(90*\angle+135:\ARROWHALFLENGTH)$) to ($(\BPSPROJ{0}{0}{\sphereradius})+(90*\angle+45:\CIRCULARMOTIONRADIUS-\ARROWHALFWIDTH)+(90*\angle+135:-\ARROWHALFLENGTH)$) to ($(\BPSPROJ{0}{0}{\sphereradius})+(90*\angle+45:\CIRCULARMOTIONRADIUS+\ARROWHALFWIDTH)+(90*\angle+135:-\ARROWHALFLENGTH)$);
	}
	\draw[bps_south_pole] (\BPSPROJ{0}{0}{-(\sphereradius)}) circle ({(#1)*(#2)*sqrt(0.5)});
	\foreach \angle in {0,...,3}{
		\fill[bps_south_pole] ($(\BPSPROJ{0}{0}{-(\sphereradius)})+(90*\angle+45:\CIRCULARMOTIONRADIUS)+(90*\angle+135:-\ARROWHALFLENGTH)$) to ($(\BPSPROJ{0}{0}{-(\sphereradius)})+(90*\angle+45:\CIRCULARMOTIONRADIUS-\ARROWHALFWIDTH)+(90*\angle+135:\ARROWHALFLENGTH)$) to ($(\BPSPROJ{0}{0}{-(\sphereradius)})+(90*\angle+45:\CIRCULARMOTIONRADIUS+\ARROWHALFWIDTH)+(90*\angle+135:\ARROWHALFLENGTH)$);
	}
	\foreach \angle in {0,...,3}{
		\draw[bps_equator] ($(\BPSPROJ{(\sphereradius)*cos(90*\angle)}{(\sphereradius)*sin(90*\angle)}{0})+(45*\angle:-\LINEARMOTIONLENGTH)$) to ($(\BPSPROJ{(\sphereradius)*cos(90*\angle)}{(\sphereradius)*sin(90*\angle)}{0})+(45*\angle:\LINEARMOTIONLENGTH)$);
	}
}

\NewDocumentCommand{\bpsDrawBlochDeco}{O{0.75}m}{
	\path[fill=white,draw=black] (\BPSPROJ{0}{0}{\sphereradius}) circle (#2);
	\path[fill=white,draw=black] (\BPSPROJ{0}{0}{-(\sphereradius)}) circle (#2);
	\draw[bps_north_pole,thick,-latex] (\BPSPROJ{0}{0}{(\sphereradius)-(#1)*(#2)}) to (\BPSPROJ{0}{0}{(\sphereradius)+(#1)*(#2)});
	\draw[bps_south_pole,thick,-latex] (\BPSPROJ{0}{0}{-(\sphereradius)+(#1)*(#2)}) to (\BPSPROJ{0}{0}{-(\sphereradius)-(#1)*(#2)});
}

\begin{document}

\title{L lines, C points and Chern numbers: understanding band structure topology using polarization fields}

\author{Thomas F\"osel\textsuperscript{1,2}, Vittorio Peano\textsuperscript{1,3} and Florian Marquardt\textsuperscript{1,2}} % TODO: order
\address{
	\textsuperscript{1} Friedrich-Alexander University Erlangen-N\"urnberg (FAU), Department of Physics, Staudtstr.\ 7, 91058 Erlangen, Germany \\
	\textsuperscript{2} Max Planck Institute for the Science of Light, Staudtstr.\ 2, 91058 Erlangen, Germany \\
	\textsuperscript{3} Department of Physics, University of Malta, Msida MSD 2080, Malta \\
% 	\textsuperscript{4} Author to whom any correspondence should be addressed.
}
% \ead{thomas.foesel@mpl.mpg.de}

\vspace{10pt}
\begin{indented}
	\item[] March 2017
\end{indented}

\begin{abstract}
	Topology has appeared in different physical contexts. The most prominent application is topologically protected edge transport in condensed matter physics. The Chern number, the topological invariant of gapped Bloch Hamiltonians, is an important quantity in this field. Another example of topology, in polarization physics, are polarization singularities, called L lines and C points. By establishing a connection between these two theories, we develop a novel technique to visualize and potentially measure the Chern number: it can be expressed either as the winding of the polarization azimuth along L lines in reciprocal space, or in terms of the handedness and the index of the C points. For mechanical systems, this is directly connected to the visible motion patterns.
\end{abstract}


\vspace{2pc}
\noindent{\it Keywords}: Chern number, polarization singularities

% \submitto{\NJP}

%\maketitle

\section{Introduction}
\label{sec:introduction}

Ever since the discovery of the transverse wave nature of electromagnetic waves, the study of the polarization properties of such vector waves has attracted a great deal of attention and led to the introduction of novel mathematical concepts. For a generic plane wave, the tip of the electric field vector traces an ellipse that defines the polarization state. In just the same way, we can describe the mechanical motion of a single pendulum that is free to move along two orthogonal directions.

For an arbitrary (monochromatic) field, its polarization becomes position-dependent, and the resulting polarizaton field can display complex spatial patterns. Again, there is a mechanical analogy, in the form of 2D arrays of coupled mechanical oscillators. Such coupled oscillator arrays have already been used as a platform to implement topologically protected transport of sound waves, using coupled pendula \cite{huber2015phononic_helical_edge_states} and coupled gyroscopes \cite{nash2015gyroscopic_metamaterials}. Eventually, they could also be realized on the nanoscale, \abbr{eg} using nanopillar arrays \cite{paulitschke2011phd,paulitschke2013gaas_nanopillars}. The time-evolution of such an array under monochromatic driving also exhibits elliptical motion that has the same mathematical description as the electromagnetic polarization fields.

The study of the complex spatial polarization patterns in random electromagnetic waves has led to interesting topological concepts. The central objects of interest are lines where the polarization gets linear, and points with circular polarization. These ``L lines'' \cite{nye1983lsurfaces,dennis2009review_polarization_sing} and ``C points'' \cite{nye1983clines,dennis2009review_polarization_sing} have been studied thoroughly in random optical fields, but are still not widely known. However, they have been found in many different physical contexts, for example the sunlight in the sky \cite{berry2004pol_sing_clear_sky} and speckle fields \cite{flossmann2008pol_sing_speckle}.

There is, of course, another branch of physics where topology has become very prominent recently: the analysis of band structures. According both to their phenomenology and their theoretical description, two categories can be identified. On the one hand, there are the Chern insulators associated with the quantum Hall effect \cite{klitzing1980qhe,klitzing1986review_qhe} and the anomalous quantum Hall effect \cite{haldane1988qhe_without_landau_levels}; they exhibit chiral edge transport as a result of a non-trivial topological invariant, the Chern number \cite{tknn1982quant_hall_conduct,avron1983homotopy_and_quantization,kohmoto1985topological_invariant}. On the other hand, topological insulators have been established by the discovery of the quantum spin Hall effect \cite{kane2005qshe,bhz2006hgte_theory,molenkamp2007hgte_experiment}; their edge channels are helical, and their topological properties are encoded in a binary ($\mathset{Z_2}$) topological invariant \cite{kane_mele2005z2_topo_order}. The importance of those topological features on transport is by now well-documented \cite{hasan_kane2010review_topo_ins}.

In the present work, it is our aim to connect these two strands of topology, in a general way that is particularly useful for mechanical systems. Our approach helps to visualize (and, in principle, measure) the Chern numbers based on the polarization fields of bulk excitations. In fact, we offer two different approaches to extract the Chern numbers of the band structure, one based on L lines and the other based on C points.

Our method is an alternative to other recently explored techniques to obtain Chern numbers for bosonic systems, both from bulk features \cite{aidelsburger2015measure_chern_ultracold_atoms,price2016measure_chern_center_of_mass}, from dynamics at the boundary \cite{mittal2016measure_chern_photonic_system} and in more general settings \cite{bardyn2014measure_chern_photonic_lattices,schroer2014measure_topo_transition}.

Besides the mechanical arrays of pendula and nanopillars mentioned above, we will also explain how our method can be applied in general to arbitrary mechanical structures (\abbr{eg} phononic crystals), and also to the electromagnetic fields in photonic structures. It is, thus, applicable in principle to a large class of the recently proposed or implemented topological devices for sound waves \cite{huber2015phononic_helical_edge_states,nash2015gyroscopic_metamaterials,prodan2009microtubules,zhang2010phonon_hall_effect,peano2015sound_and_light,wang2015topo_phononic_crystal,yang2015topo_acoustics,kariyado2015mech_graphene,chen2016tunable_topo_phononic_crystal,fleury2016floquet_topo_ins,matlack2016designing,mitchell2016amorphous_gyroscopic,souslov2016topological_liquid}, light waves \cite{lu2014review_topo_photonics,haldane2008possible_in_photonic_crystals,raghu2008qhe_analogs_in_photonic_crystals,wang2009topo_elm_states,koch2010time_reversal_symm_breaking_in_circuit_qed,umucalilar2011artificial_gauge_field,hafezi2011robust_optical_delay_lines,fang2012effective_magn_field,petrescu2012ahe_light_and_chiral_kagome,hafezi2013imaging_topo_edge_states,schmidt2015optomech_magn_fields,peano2016topological_quantum_amplifier,anderson2016engineering_topo_in_microwave_cavity_arrays}, ultracold atoms \cite{jotzu2014haldane_model_ultracold_fermions,goldman2014review_ultracold_fermions}, or magnons \cite{onose2010review_magnon_hall_effect,shindou2013topological_magnonic_crystal,zhang2013topo_magnon_insulator,mook2014topo_magnon_insulators,chisnell2015topo_magnon_bands}. These include Chern insulators, as well as time-reversal preserving topological insulators whose Hamiltonian can be decomposed into a pair of Chern insulator Hamiltonians with opposite Chern numbers.

\section{Connecting topological bands to polarization fields}
\label{sec:preliminaries}

In the following, we will consider a particle or excitation hopping around on a lattice in two dimensions. We will assume that there is some internal degree of freedom associated with the particle, or, equivalently, a certain number of basis states situated at each lattice site. The physical origin may vary: the spin of a particle, the atomic orbital for atoms in a crystal (s, p, d, \dots), the different oscillation directions of a mechanical resonator in a phononic lattice, the sense of circulation of the whispering gallery modes in a lattice of coupled optical disk resonators\cite{hafezi2012microring_resonators}, or the polarization of photon-exciton polaritons in micropillars\cite{jacqmin2014honeycomb_polaritons,sala2015soc_polaritons,nalitov2015soc_photonic_graphene}. In addition, one unit cell might consist of multiple sites, depending on the lattice geometry. For all these cases, we will speak of the ``polarization degree of freedom''.

The total number $N$ of basis states associated with a unit cell of the Bravais lattice determines the number of bands. In the following, we will denote the real-space basis states as $\ket{j,s}$ where $j\in\mathset{Z2}$ labels the unit cell and $s\in\{1,\hdots,N\}$ refers to the polarization degree of freedom.

The eigenmodes of every Hamiltonian $\hat{H}$ which respects the translational invariance are the Bloch states
\begin{equation}
	\ket{\Phi_n(\vec{k})} = \sum_{j,s} \Psi_s^{(n)}(\vec{k})\,\mconst{e}^{i\vec{k}\cdot\vec{r}_j} \ket{j,s}
	\label{eq:definition_bloch_state}
\end{equation}
where $n$ is the index of the band (with $1\leq n\leq N$) and $\vec{r}_j$ the position of the unit cell.

We now consider the Chern number of the band $n$:
\begin{equation}
	C_n = \frac{1}{2\mconst{pi}} \iint_\mathrm{BZ} \Big[\vec{\nabla}_{\vec{k}}\times \vec{\cal{A}}_n(\vec{k})\Big]\cdot \vec{\mathrm{e}}_z\,\total[^2]{k} \in \mathset{Z}
	\label{eq:definition_chern_number}
\end{equation}
where $\vec{\cal{A}}_n(\vec{k})=-\mathrm{i}\,\Braket{\Phi_n(\vec{k})}{\vec{\nabla}_{\vec{k}}}{\Phi_n(\vec{k})}$ is the Berry connection and $\vec{\mathrm{e}}_z$ is the unit vector in the direction of the $z$-axis. The complex amplitudes $\Psi_s^{(n)}(\vec{k})$ completely characterize the Bloch states in the band $n$. They can already be deduced by inspecting a Bloch state in a single unit cell. Observation of the polarization pattern within one unit cell as a function of $\vec{k}$ is therefore sufficient to determine the Chern number of the $n$-th band. This may even be of experimental relevance, for platforms like mechanical systems where direct measurement of the motion pattern in a given Bloch state is feasible.

It turns out that the Chern number can even be calculated just by knowing the projections of the Bloch state $\ket{\Phi_n(\vec{k})}$ along two linearly independent directions, as long as these projections do not both vanish simultaneously for the same quasimomentum $\vec{k}$ (for a proof, see \cref{sec:connection+proof}). Without loss of generality, we choose orthogonal directions in our polarization basis. The two projections define a complex vector field in reciprocal space:
\begin{equation}
	\vec{\psi}^{(n)}(\vec{k}) =
	\begin{pmatrix}
		\Psi_{s_1}^{(n)}(\vec{k}) \\
		\Psi_{s_2}^{(n)}(\vec{k})
	\end{pmatrix}
	\in \mathset{C2}
	\label{eq:definition_polarized_field}
\end{equation}
For particles with spin-1/2 or equivalent systems like mechanical oscillators with two degrees of freedom (oscillation directions), $s_1$ and $s_2$ can refer to the two basis states of the internal degree of freedom. If the lattice is a non-Bravais lattice, the projection onto one sublattice is a natural choice. In the mechanical case, $\vec{\psi}^{(n)}(\vec{k})$ corresponds to the directly visible motion pattern $\vec{r}_{\vec{k}}(t)=\Real{\vec{\psi}^{(n)}(\vec{k})\mconst{e}^{-\mconst{i}\omega t}}$ on the selected site. We note that, in the most general case, the orbit will be an ellipse.

\subsection{Geometrical interpretation of the polarization}

\begin{figure}[b!]
	\centering
	\begin{tikzpicture}
		\pgfmathsetmacro{\scale}{0.8}
		\pgfmathsetmacro{\widthAB}{9.5*\scale}
		\pgfmathsetmacro{\widthC}{9*\scale}
		\pgfmathsetmacro{\heightA}{6*\scale}
		\pgfmathsetmacro{\heightB}{3*\scale}
		\pgfmathsetmacro{\margin}{0.5*\scale}
		
		\pgfmathsetmacro{\imgBscale}{0.8*\scale}
		
		\coordinate (a_nw) at (0,\heightB+\margin+\heightA);
		\coordinate (a_se) at (\widthAB,\heightB+\margin);
		\coordinate (b_nw) at (0,\heightB);
		\coordinate (b_se) at (\widthAB,0);
		\coordinate (c_nw) at (\widthAB+\margin,\heightB+\margin+\heightA);
		\coordinate (c_se) at (\widthAB+\margin+\widthC,0);
		
		% subfigure a
		\begin{scope}[shift={($(a_nw)!0.5!(a_se)$)},declare function={normalisation(\majorAxis,\minorAxis,\phi)=sqrt(2/(\majorAxis^2+\minorAxis^2-(\majorAxis^2-\minorAxis^2)*cos(2*\phi)));}]
			\pgfmathsetmacro{\paramNorm}{4*\scale}
			\pgfmathsetmacro{\paramPhi}{35}
			\pgfmathsetmacro{\paramTheta}{25}
			\pgfmathsetmacro{\paramZeta}{17}
			
			\pgfmathsetmacro{\thetaArcSize}{1.5*\scale}
			\pgfmathsetmacro{\phiArcDistance}{0.2*\scale}
			\pgfmathsetmacro{\lengthAxisX}{4.5*\scale}
			\pgfmathsetmacro{\lengthAxisY}{2.75*\scale}
			\pgfmathsetmacro{\gapUp}{0.3}
			\pgfmathsetmacro{\gapDown}{0.2}
			
			\pgfmathsetmacro{\majorAxis}{\paramNorm*cos(\paramZeta)}
			\pgfmathsetmacro{\minorAxis}{\paramNorm*sin(\paramZeta)}
			
			% draw the coordinate axes; the x axis is discontinued for the label of the minor semi-axis of the ellipse
			\draw[white!50!black,line width=0.8,-latex] (-\lengthAxisX,0) -- (\lengthAxisX,0) node[below left] {x};
			\draw[white!50!black,line width=0.8,-latex] (0,-\lengthAxisY) -- (0,\lengthAxisY) node[below right] {y};
			
			% draw the minor and major semi-axis of the ellipse
			\draw[ForestGreen] (0,0) -- node[above,rotate=\paramTheta] {$\sqrt{\pquant{S0}}\cdot\cos\pquant{polar_altitude}$} (\paramTheta:\majorAxis);
			\draw[ForestGreen] (0,0) -- node[left,rotate=\paramTheta,fill=white,fill opacity=0.5,text opacity=1] {$\sqrt{\pquant{S0}}\cdot\sin\pquant{polar_altitude}$} (\paramTheta+90:\minorAxis);
			
			% draw the angle theta
			\draw[red] (\thetaArcSize,0) arc (0:\paramTheta:\thetaArcSize);
			\node[red] at (0.5*\paramTheta:\thetaArcSize-0.3) {$\pquant{polar_azimuth}$};
			
			% draw the ellipse itself
			\draw[blue,thick,rotate=\paramTheta] (0,0) ellipse ({\majorAxis} and {\minorAxis});
			
			% tag the point t=0
			\coordinate (time_zero) at ({\majorAxis*cos(\paramTheta)*cos(\paramPhi)+\minorAxis*sin(\paramTheta)*sin(\paramPhi)},{\majorAxis*sin(\paramTheta)*cos(\paramPhi)-\minorAxis*cos(\paramTheta)*sin(\paramPhi)});
			\fill[Black] (time_zero) circle (0.08);
			\node at ($(time_zero)+(\paramTheta-85:0.4)$) {$t=0$};
			
			% visualize the phase angle phi
			\draw[Magenta,variable=\phi,domain=-\paramPhi:0,rotate=\paramTheta,<->] plot ({(\majorAxis+normalisation(\majorAxis,\minorAxis,\phi)*\phiArcDistance*\minorAxis)*cos(\phi)},{(\minorAxis+normalisation(\majorAxis,\minorAxis,\phi)*\phiArcDistance*\majorAxis)*sin(\phi)}) node[below right=5] {$\omega\Delta t=\varphi$};
		\end{scope}
		
		% subfigure b
		\node at ($(b_nw)!0.5!(b_se)$) {\includegraphics[scale={\imgBscale}]{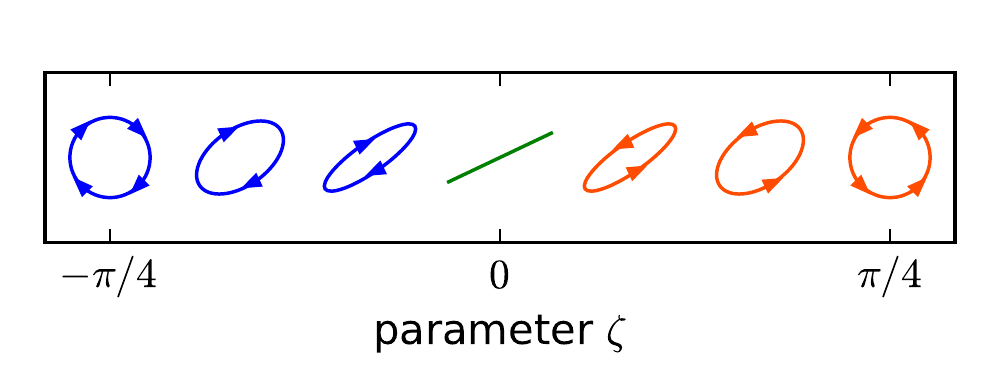}};
		
		% subfigure c
		\begin{scope}[shift={($(c_nw)!0.5!(c_se)+(0,-0.75*\scale)$)}]
			\pgfmathsetmacro{\sphereradius}{4*\scale}
			\pgfmathsetmacro{\azimut}{223}
			\pgfmathsetmacro{\elevation}{29}
			
			\bpsDrawSphere{shaded}
			\bpsDrawAxes[poincare][0.3]{1.4}
			\bpsDrawAngles[$2\pquant{polar_azimuth}$][-0.5]{35}[$2\pquant{polar_altitude}$][0.8][0.6]{20}{0.55}
			\bpsDrawPoincareDeco{0.4*\scale}{0.08}{0.1}
		\end{scope}
		
		% subfigure labels
		\node[anchor=north west] at (a_nw) {a)};
		\node[anchor=north west] at (b_nw) {b)};
		\node[anchor=north west] at (c_nw) {c)};
		
% 		% subfigure frames
% 		\draw[cyan] (a_nw) rectangle (a_se);
% 		\draw[cyan] (b_nw) rectangle (b_se);
% 		\draw[cyan] (c_nw) rectangle (c_se);
	\end{tikzpicture}
	\caption{(a) Parameterization of an elliptical motion pattern: $\pquant{S0}$ determines the scaling, $\pquant{polar_azimuth}$ is the rotation angle of the main axes, and $\varphi$ corresponds to a time offset. (b) Motion patterns for various values of the polarization altitude $\pquant{polar_altitude}$ which determines the handedness and the ellipticity, ranging from circular right-handed (south pole, $\pquant{polar_altitude}=-\frac{1}{4}\mconst{pi}$) over linear (equator, $\pquant{polar_altitude}=0$) to circular left-handed polarization (north pole, $\pquant{polar_altitude}=\frac{1}{4}\mconst{pi}$). (c) Representation on the \name{Poincare} sphere. The three coordinate axes $\pquant{S1}$, $\pquant{S2}$ and $\pquant{S3}$ are the Stokes parameters. A half rotation of the major axis in the two-dimensional physical space is already sufficient for a full rotation around the $\pquant{S3}$ axis since it only gives rise to a global minus sign which can be absorbed into the complex phase $\varphi$. Note that two orthogonal states ($\braket{a}{b}=0$) are located on opposite spots of the \name{Poincare} sphere.}
	\label{fig:polarization+poincare_sphere}
\end{figure}
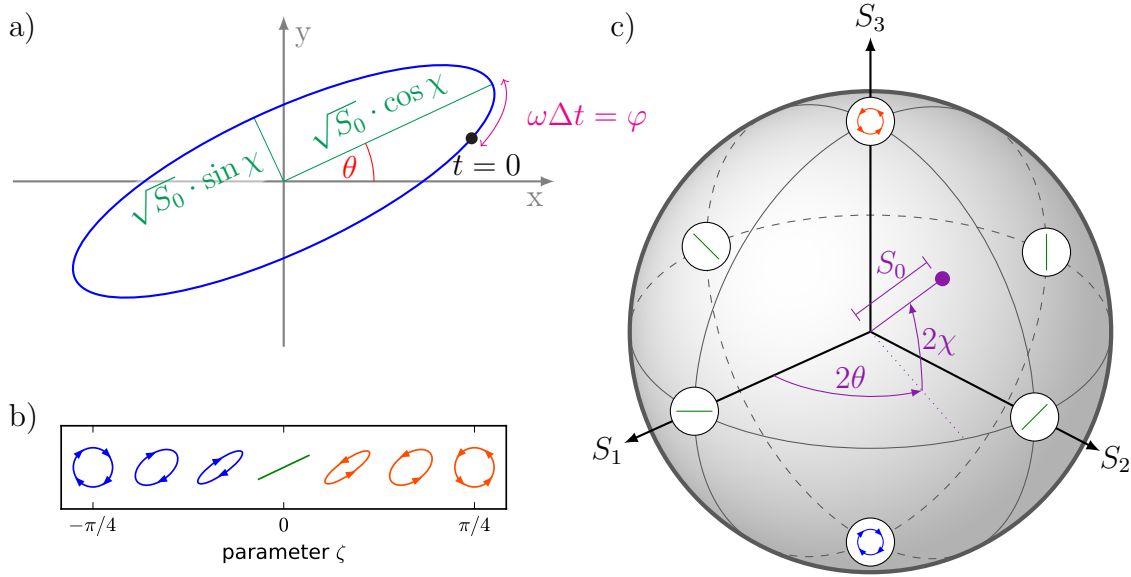

The vector field $\vec{\psi}^{(n)}(\vec{k})$ defined in \cref{eq:definition_polarized_field} can be rewritten in the useful parameterization
\begin{equation}
	\vec{\psi}^{(n)}(\vec{k}) = \sqrt{\pquant{S0}} \mconst{e}^{\mconst{i}\varphi} \begin{pmatrix}\cos\pquant{polar_azimuth}&-\sin\pquant{polar_azimuth}\\\sin\pquant{polar_azimuth}&\cos\pquant{polar_azimuth}\end{pmatrix} \begin{pmatrix}\cos\pquant{polar_altitude}\\\mconst{i}\sin\pquant{polar_altitude}\end{pmatrix}.
	\label{eq:parameterization_poincare}
\end{equation}
In a mechanical setting, the angle $\pquant{polar_azimuth}(\vec{k})$ gives the direction of the major axis of the elliptical orbit and is therefore called the polarization azimuth, $\pquant{polar_altitude}(\vec{k})$ represents a measure of its ellipticity and handnedness, and $\pquant{S0}$ (the total intensity along the two projections) is the squared diagonal of the axis-aligned bounding box, \abbr{cf} \cref{fig:polarization+poincare_sphere} (a-b). In addition, $\varphi$ is the oscillation phase which is a gauge of freedom for the eigenstate.

One can characterize the polarization of the field $\vec{\psi}^{(n)}(\vec{k})$ by the Stokes parameters\cite{born_wolf1959stokes_params,brosseau1998fundamentals}
\begin{subequations}
	\begin{align}
		\pquant{S0} & = \abs{\psi_1}^2+\abs{\psi_2}^2 \\
		\pquant{S1} & = \abs{\psi_1}^2-\abs{\psi_2}^2 = \pquant{S0} \cos(2\pquant{polar_altitude}) \cos(2\pquant{polar_azimuth}) \\
		\pquant{S2} & = 2\Real{\conj{\psi_1}\psi_2}   = \pquant{S0} \cos(2\pquant{polar_altitude}) \sin(2\pquant{polar_azimuth}) \\
		\pquant{S3} & = 2\Imag{\conj{\psi_1}\psi_2}   = \pquant{S0} \sin(2\pquant{polar_altitude}).
	\end{align}
	\label{eq:definition_stokes_parameters}
\end{subequations}
The parameters $S_1$, $S_2$, and $S_3$ can be identified with the coordinates on the \name{Poincare} sphere, while $\pquant{S0}$ is the corresponding radius, \abbr{cf} \cref{fig:polarization+poincare_sphere} (c).

\section{L Lines and Chern numbers}
\label{sec:l_lines}

In the previous section, we have shown how to assign to each band $n$ an auxiliary polarization field $\vec{\psi}^{(n)}(\vec{k})$, see \cref{eq:definition_polarized_field}. In the remainder of the paper, we will make use of this field and of the geometrical interpretation of the polarization to provide a recipe to visualize the Chern number of the band.

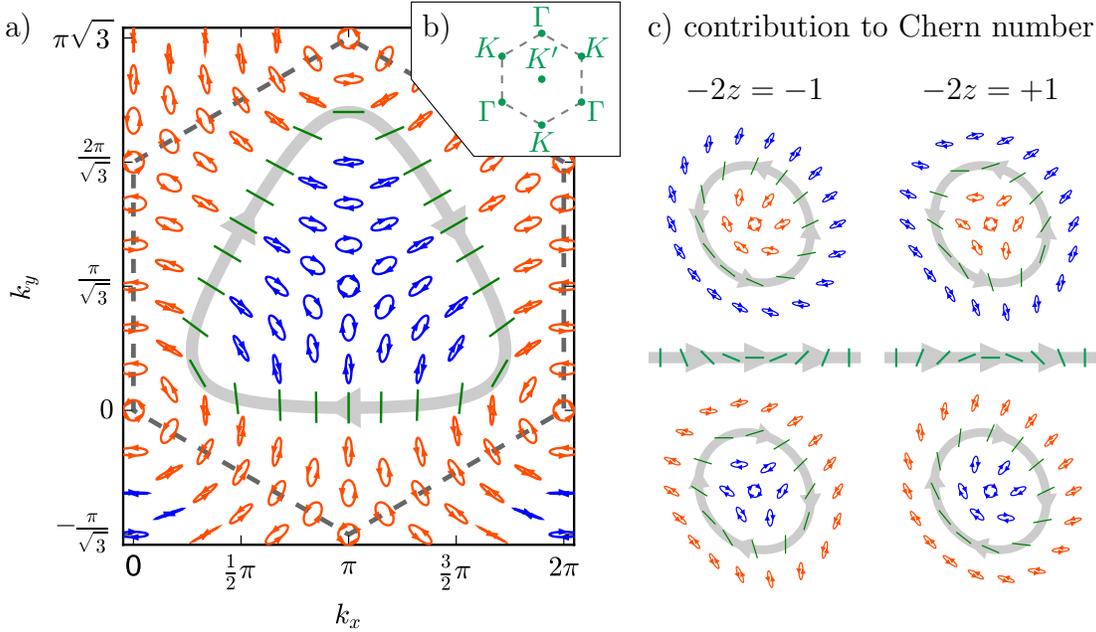
\begin{figure}[t!]
	\centering
	\begin{tikzpicture}[subfig/.style 2 args={inner sep=0,fit=(#1)(#2)}]
		\pgfmathsetmacro{\scale}{1.11}
		\pgfmathsetmacro{\widthA}{7.5*\scale}
		\pgfmathsetmacro{\widthB}{2.5*\scale}
		\pgfmathsetmacro{\widthC}{5.65*\scale}
		\pgfmathsetmacro{\height}{7.75*\scale}
		\pgfmathsetmacro{\heightB}{1.85*\scale}
		\pgfmathsetmacro{\margin}{0.2*\scale}
		
		\pgfmathsetmacro{\bzRadius}{0.55*\scale}
		\pgfmathsetmacro{\scaleImgA}{0.8*\scale}
		\pgfmathsetmacro{\scaleImgC}{0.42*\scale}
		
		\coordinate (a_nw) at (0,\height);
		\coordinate (a_se) at (\widthA,0);
		\coordinate (b_nw) at (\widthA-\widthB,\height);
		\coordinate (b_se) at (\widthA,\height-\heightB);
		\coordinate (c_nw) at (\widthA+\margin,\height);
		\coordinate (c_se) at (\widthA+\margin+\widthC,0);
		\coordinate (b_bz) at (\widthA-0.375*\widthB,\height-0.5*\heightB);
		\coordinate (c_negative) at (\widthA+\margin+0.25*\widthC,0.45*\height);
		\coordinate (c_positive) at (\widthA+\margin+0.75*\widthC,0.45*\height);
		\coordinate (c_schematic_offset) at (0,1.6*\scale);
		\coordinate (c_label_offset) at (0,3.2*\scale);
		
		% subfigure a)
		\node at ($(a_nw)!0.5!(a_se)$) {\includegraphics[scale=\scaleImgA]{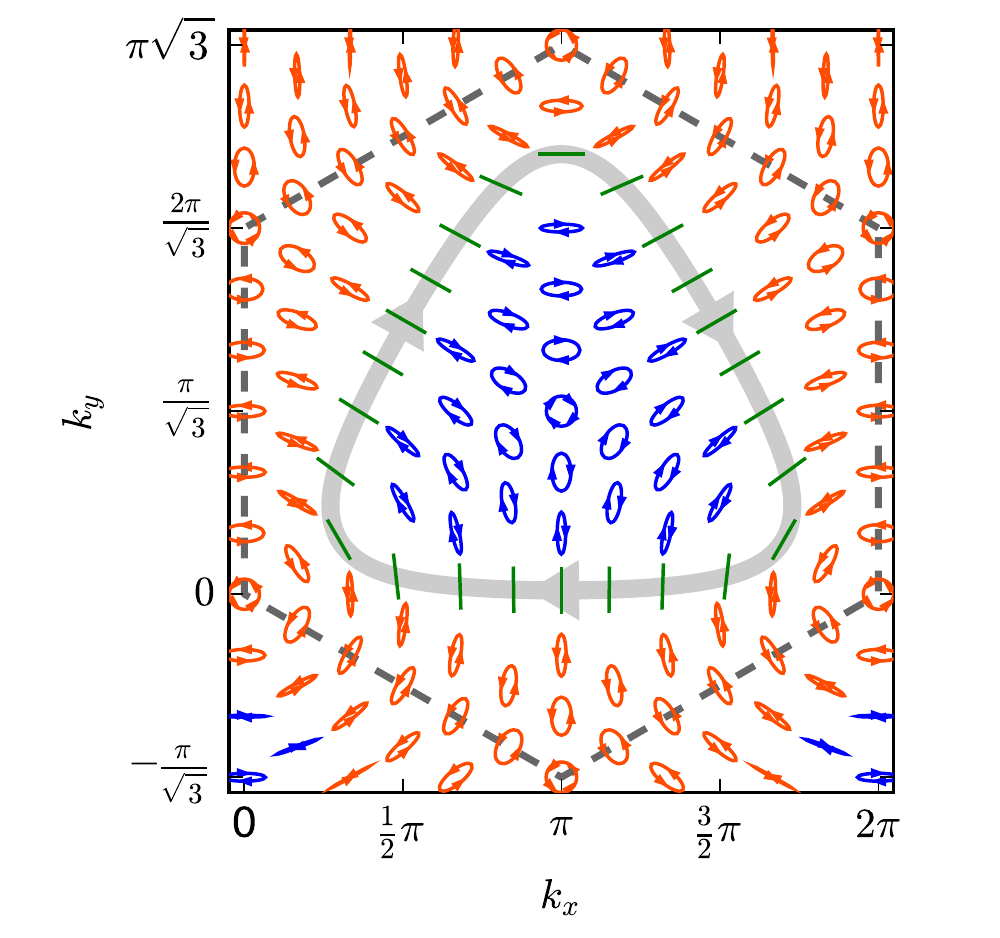}};
		
		% subfigure b)
		\path[draw=black,fill=white] (b_nw) to (b_nw-|b_se) to (b_se) to ($(b_se-|b_nw)!0.3!(b_se)$) to ($(b_se-|b_nw)!0.5!(b_nw)$) -- cycle;
		\draw[gray,thick,dashed] ($(b_bz)+(30:\bzRadius)$) \foreach \angle in {1,...,5} {to ($(b_bz)+(30+60*\angle:\bzRadius)$)} -- cycle;
% 		\foreach \angle in {0,...,2}{
% 			\draw[gray,thick,dashed] (b_bz) to ($(b_bz)+(120*\angle+30:\bzRadius)$);
% 			\draw[gray,thick,dashed] ($(b_bz)+(120*\angle+30:\bzRadius)+(120*\angle+90:0.75*\bzRadius)$) to ($(b_bz)+(120*\angle+30:\bzRadius)$) to ($(b_bz)+(120*\angle+30:\bzRadius)+(120*\angle-30:0.75*\bzRadius)$);
% 		}
		\fill[ForestGreen] (b_bz) circle (0.05);
		\foreach \angle in {0,...,5}{
			\fill[ForestGreen] ($(b_bz)+(60*\angle+30:\bzRadius)$) circle (0.05);
		}
		\node[above,ForestGreen] at (b_bz) {\small{$K'$}};
		\foreach \angle in {0,...,2}{
			\node[ForestGreen] at ($(b_bz)+(120*\angle+30:\bzRadius+0.225)$) {\small{$K$}};
			\node[ForestGreen] at ($(b_bz)+(120*\angle-30:\bzRadius+0.225)$) {\small{$\Gamma$}};
		}
		
		% subfigure c)
		\node at ($(c_negative)+(c_schematic_offset)$)  {\includegraphics[scale=\scaleImgC]{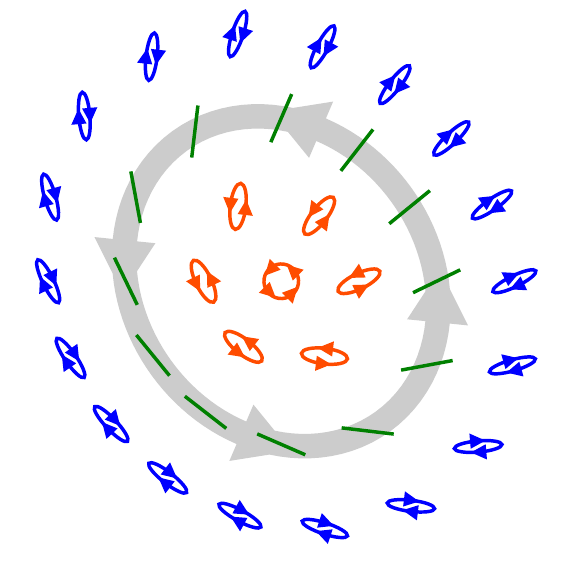}};
		\node at ($(c_negative)-(c_schematic_offset)$)  {\includegraphics[scale=\scaleImgC]{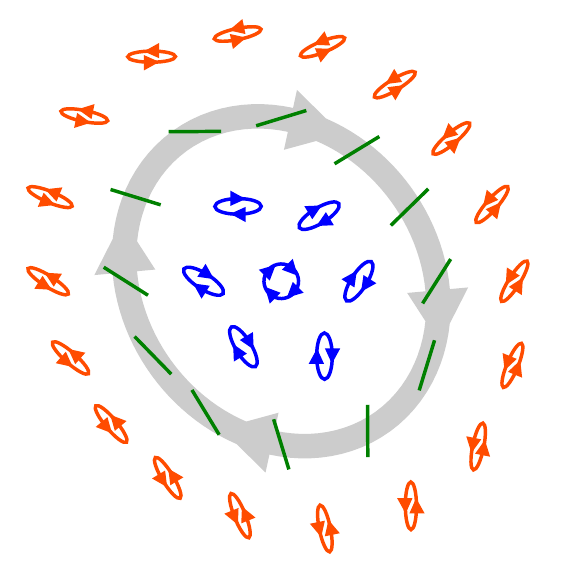}};
		\node at ($(c_negative)+(c_label_offset)$) {$-2z=-1$};
		\multiarrow[fill=black!20!white,shift=(c_negative)]{0}{0}{0.45*\widthC}{0.1}{0.15}{0.4}{3}
		\winding[ForestGreen,thick,shift=(c_negative)]{0}{0}{0.4*\widthC}{0.25}{8}{1} % note that the sign of the winding is opposite to the contribution to the Chern number
		\node at ($(c_positive)+(c_schematic_offset)$) {\includegraphics[scale=\scaleImgC]{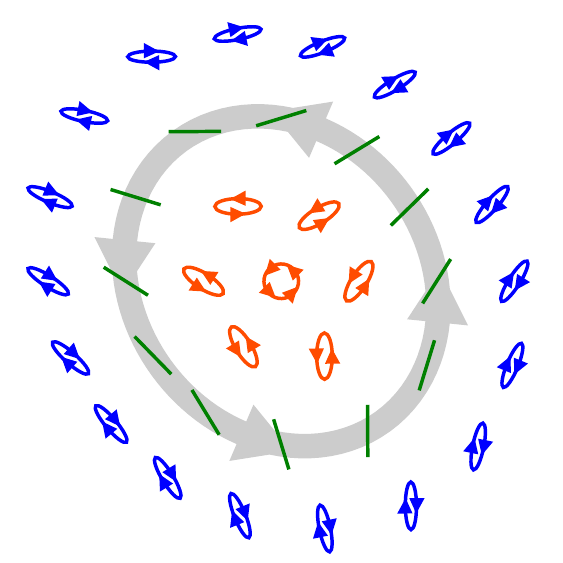}};
		\node at ($(c_positive)-(c_schematic_offset)$) {\includegraphics[scale=\scaleImgC]{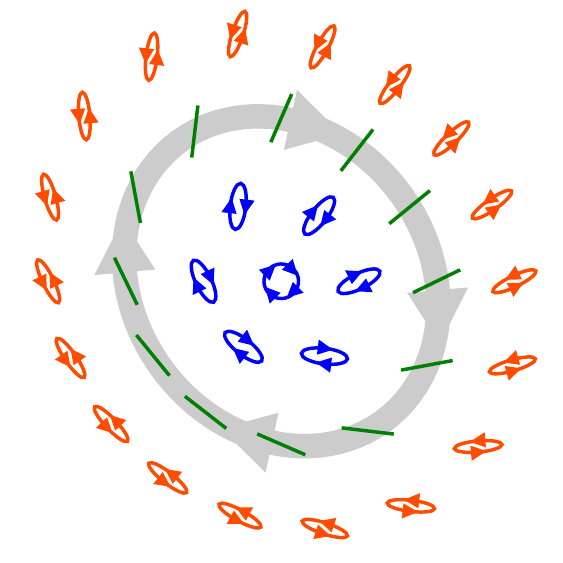}};
		\node at ($(c_positive)+(c_label_offset)$) {$-2z=+1$};
		\multiarrow[fill=black!20!white,shift=(c_positive)]{0}{0}{0.45*\widthC}{0.1}{0.15}{0.4}{3}
		\winding[ForestGreen,thick,shift=(c_positive)]{0}{0}{0.4*\widthC}{0.25}{8}{-1} % note that the sign of the winding is opposite to the contribution to the Chern number
		
		% subfigure labels
		\node[anchor=north west] at (a_nw) {a)};
		\node[anchor=north west] at (b_nw) {b)};
		\node[anchor=north west] at (c_nw) {c) contribution to Chern number};
		
% 		% subfigure frames
% 		\draw[draw=cyan] (a_nw) rectangle (a_se);
% 		\draw[draw=cyan] (b_nw) rectangle (b_se);
% 		\draw[draw=cyan] (c_nw) rectangle (c_se);
	\end{tikzpicture}
	\caption{Interpretation of the Chern number in terms of L lines.
		(a) shows the polarization pattern $\vec{\psi}^{(1)}(\vec{k})$ for the lowest-frequency band of the mechanical model described in \cref{sec:application} ($t_\mathrm{T}/t_\mathrm{L}=0.2$, $t_\mathrm{Zeeman}/t_\mathrm{L}=0.02$). The polarization is right-handed in an area centered around the $K'$ point, and left-handed in the other part of the dashed hexagon. We have indicated the L line, \abbr{ie} the boundary of these regions where the polarization is linear. While traversing the L line in clockwise direction, the polarization azimuth $\theta$ winds up by $\mconst{pi}$, so $z=\frac{1}{2}$. As explained in \cref{sec:l_lines}, this can be used to predict the Chern number, which is $-2z=-1$. In order to make the L line better visible, the plot does not show the basic Brillouin zone itself, but a shifted version as indicated in (b).
		(c) illustrates the general recipe: the traversal direction along the L line is determined by the handedness of the polarization in the enclosed region, \abbr{ie} counter-clockwise (clockwise) in two upper (lower) cases; the winding of the polarization azimuth $\pquant{polar_azimuth}$ is $2\mconst{pi}z$. Any $z\in\frac{1}{2}\mathset{Z}$ is allowed, and the four examples display all the possibilities with $z=\pm\frac{1}{2}$. From the winding numbers $z_l$ for all the L lines in the Brillouin zone, the Chern number $C_n$ can be obtained according to \cref{eq:relation_chern+l_lines}.
	}
	\label{fig:l_lines+chern}
\end{figure}

A typical polarization pattern is shown in \cref{fig:l_lines+chern}\nolinebreak[4](a). There, it can be seen that the polarization gets perfectly linear at the interfaces between the regions with left- and right-handed polarization. These interfaces are therefore called L lines. They are a generic feature of continuous polarized fields. L lines have been studied especially in polarized random fields \cite{nye1983lsurfaces,dennis2009review_polarization_sing}. They are structurally stable upon small perturbations $\vec{\psi}(\vec{k})\to\vec{\psi}(\vec{k})+\delta\vec{\psi}(\vec{k})$. Their robustness can be motivated by a topological argument: because the mapping on the \name{Poincare} sphere is continuous, any path between two points on different hemispheres, \abbr{ie} opposite handedness, has to inevitably cross the equator where the polarization is linear.

Below, we explain how to simply read off the Chern numbers by inspecting the L lines of the polarization field. This is possible under a basic assumption: there should be no amplitude vortex in the polarization field. Amplitude vortices are the points where $\pquant{S0}$ vanishes. In two dimensions, they are not topologically protected because they require that four parameters vanish simultaneously (the real and imaginary part of $\psi_1$ and $\psi_2$). Thus, they do not appear in generic random fields. While they could emerge as a consequence of some lattice point symmetry, they will disappear in the presence of a small perturbation that breaks that symmetry (and does not change the Chern number). Even without modifying the underlying Hamiltonian, one could eliminate such structures by a different choice of the projections used to define the auxiliary polarization field $\vec{\psi}^{(n)}(\vec{k})$. For the above reasons, the scenario analyzed here where no amplitude vortex is present is not a special case but rather a generic one.

In view of establishing a connection between the L lines and the Chern numbers, we define the winding number $z$ of the polarization azimuth along one L line. This is defined as follows: We observe how the linear polarization direction changes as we traverse the closed L line in a specified sense. If the polarization turns around the origin counterclockwise (clockwise), $z$ acquires a positive (negative) sign, and its absolute value is determined by the number of full turns. This value can be half-integer, since the linear polarization direction is defined only modulo 180 degrees. For our purposes, we specify that the traversal direction of the L line should always be chosen to match the handedness of the polarization field in the enclosed region (clockwise in the example of \cref{fig:l_lines+chern}\nolinebreak[4]\,a).

We now state (and later prove) one of the main messages of this paper: The winding number $z$ of the polarization azimuth along this L line is directly connected to the Chern number. In the simplest case with one L line per Brillouin zone, it is directly given by $C_n=-2z$. If there are multiple L lines per Brillouin zone, their contributions simply add up:
\begin{equation}
	C_n = -2 \sum_{l\in\mathcal{L}_n} z_l
	\label{eq:relation_chern+l_lines}
\end{equation}
In particular, the Chern number is automatically $0$ if there are no L lines at all.

\section{C Point classifications and Chern numbers}
\label{sec:c_points}

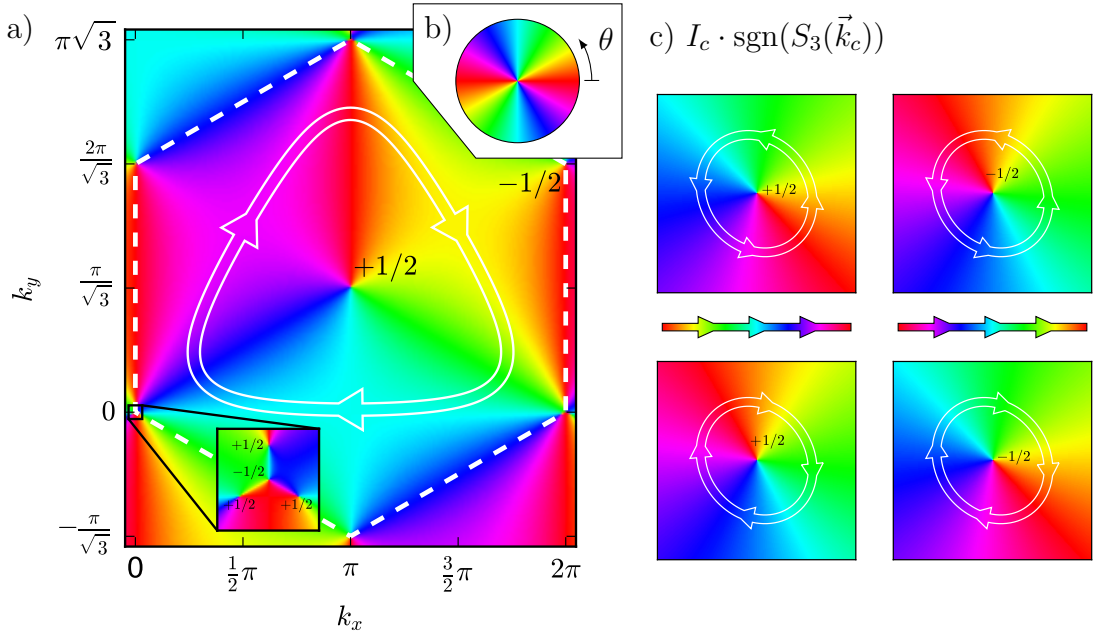
\begin{figure}[b!]
	\centering
	\begin{tikzpicture}[subfig/.style 2 args={inner sep=0,fit=(#1)(#2)}]
		\pgfmathsetmacro{\scale}{1.11}
		\pgfmathsetmacro{\widthA}{7.5*\scale}
		\pgfmathsetmacro{\widthB}{2.5*\scale}
		\pgfmathsetmacro{\widthC}{5.65*\scale}
		\pgfmathsetmacro{\height}{7.75*\scale}
		\pgfmathsetmacro{\heightB}{1.85*\scale}
		\pgfmathsetmacro{\margin}{0.2*\scale}
		
		\pgfmathsetmacro{\bzRadius}{0.55*\scale}
		\pgfmathsetmacro{\scaleImgA}{0.8*\scale}
		\pgfmathsetmacro{\scaleImgC}{0.42*\scale}
		
		\coordinate (a_nw) at (0,\height);
		\coordinate (a_se) at (\widthA,0);
		\coordinate (b_nw) at (\widthA-\widthB,\height);
		\coordinate (b_se) at (\widthA,\height-\heightB);
		\coordinate (c_nw) at (\widthA+\margin,\height);
		\coordinate (c_se) at (\widthA+\margin+\widthC,0);
		\coordinate (b_dcw) at (\widthA-0.5*\widthB,\height-0.5*\heightB);
		\coordinate (c_negative) at (\widthA+\margin+0.25*\widthC,0.5*\height);
		\coordinate (c_positive) at (\widthA+\margin+0.75*\widthC,0.5*\height);
		\coordinate (c_schematic_offset) at (0,1.6*\scale);
		
		% subfigure a)
		\node at ($(a_nw)!0.5!(a_se)$) {\includegraphics[scale=\scaleImgA]{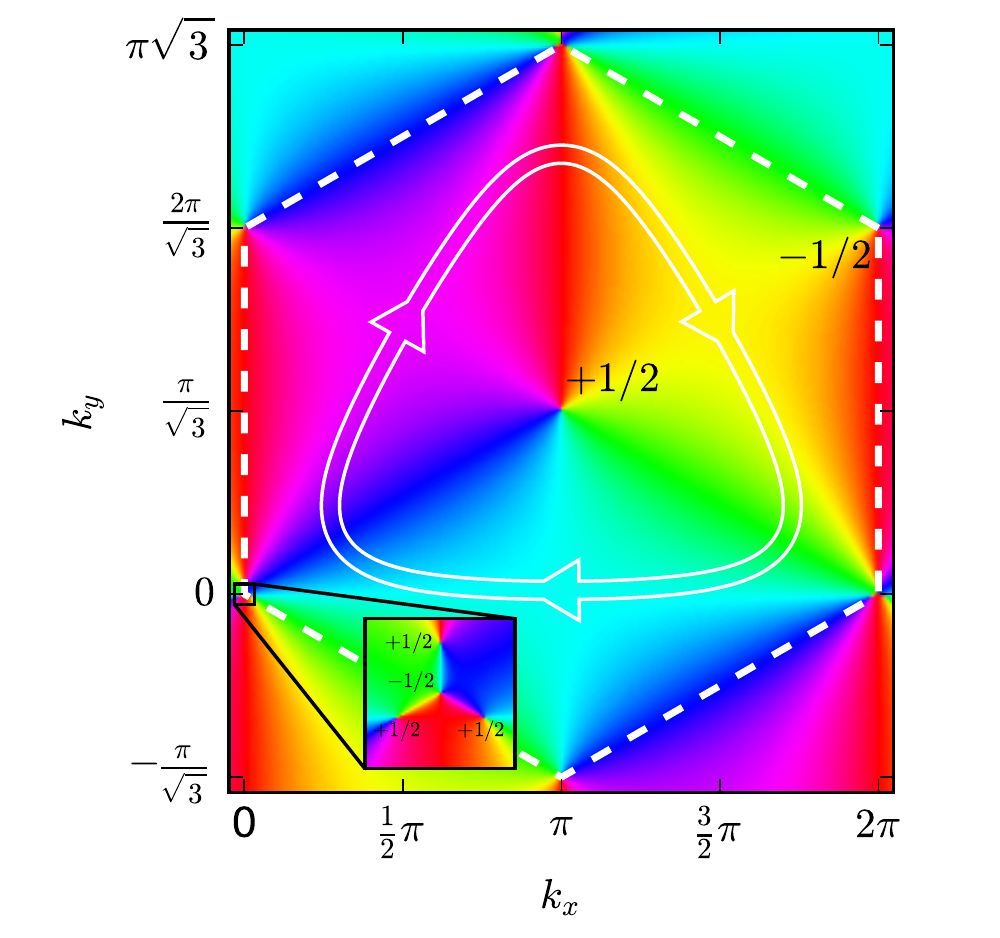}};
		
		% subfigure b)
		\path[draw=black,fill=white] (b_nw) to (b_nw-|b_se) to (b_se) to ($(b_se-|b_nw)!0.3!(b_se)$) to ($(b_se-|b_nw)!0.5!(b_nw)$) -- cycle;
		\path[draw=black,fill stretch image=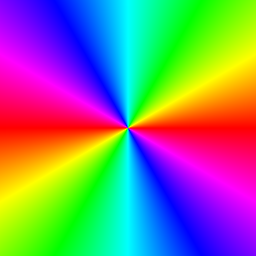] (b_dcw) circle (0.4*\heightB);
		\draw[|-latex] ($(b_dcw)+(0:0.48*\heightB)$) arc (0:35:0.48*\heightB) node[right=0.3em] {$\pquant{polar_azimuth}$};
		
		% subfigure c)
		\node at ($(c_negative)+(c_schematic_offset)$) {\includegraphics[scale=\scaleImgC]{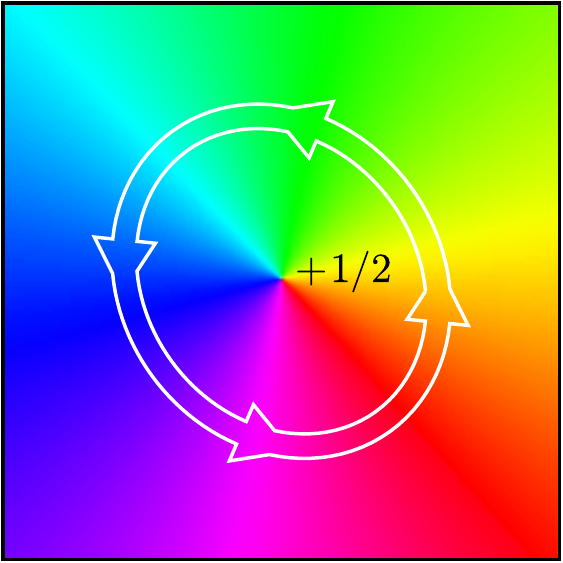}};
		\node at ($(c_negative)-(c_schematic_offset)$) {\includegraphics[scale=\scaleImgC]{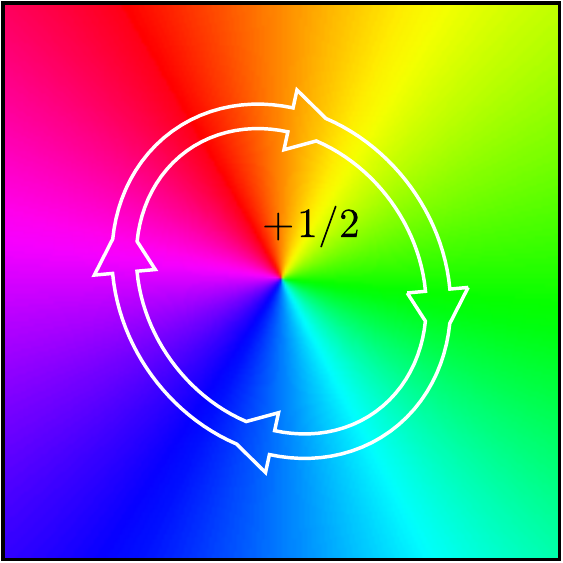}};
		\multiarrow[draw=black,fill stretch image=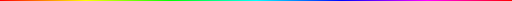,shift=(c_negative)]{0}{0}{0.4*\widthC}{0.08}{0.016*\widthC}{0.045*\widthC}{3}
		\node at ($(c_positive)+(c_schematic_offset)$) {\includegraphics[scale=\scaleImgC]{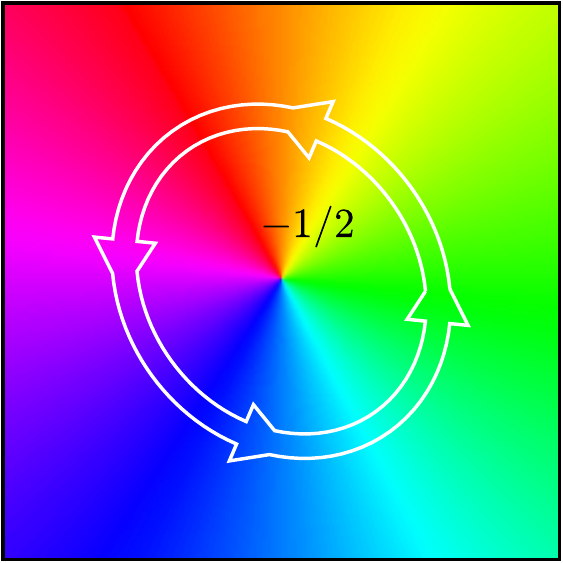}};
		\node at ($(c_positive)-(c_schematic_offset)$) {\includegraphics[scale=\scaleImgC]{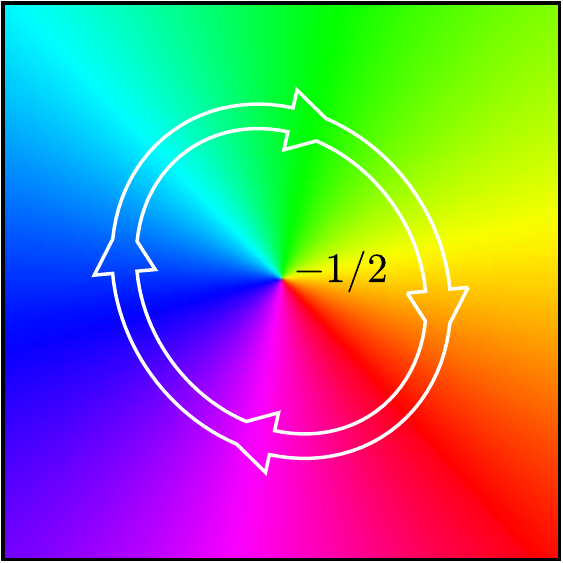}};
		\multiarrow[draw=black,fill stretch image*={angle=180,origin=c}{img/hsv.png},shift=(c_positive)]{0}{0}{0.4*\widthC}{0.08}{0.016*\widthC}{0.045*\widthC}{3}
		
		% subfigure labels
		\node[anchor=north west] at (a_nw) {a)};
		\node[anchor=north west] at (b_nw) {b)};
		\node[anchor=north west] at (c_nw) {c) $I_c \cdot \mathrm{sgn}(S_3(\vec{k}_c))$};
		
% 		% subfigure frames
% 		\draw[draw=cyan] (a_nw) rectangle (a_se);
% 		\draw[draw=cyan] (b_nw) rectangle (b_se);
% 		\draw[draw=cyan] (c_nw) rectangle (c_se);
	\end{tikzpicture}
	\caption{Interpretation of the Chern number in terms of C points.
		(a) plots the polarization azimuth $\pquant{polar_azimuth}$, \abbr{ie} how the major axes of the polarization patterns are oriented, for the same field $\vec{\psi}^{(1)}(\vec{k})$ as in \cref{fig:l_lines+chern}\,a. The color wheel in (b) tells how to translate a color into the corresponding direction; note that $\pquant{polar_azimuth}$ is defined only up to $\mconst{pi}$ (not $2\mconst{pi}$). We can see that the $\pquant{polar_azimuth}$ field in (a) has six singularities (vortices) per Brillouin zone; they correspond to the C points in $\vec{\psi}^{(1)}(\vec{k})$. The L line, drawn in white (with arrows indicating the handedness of the polarization field enclosed inside the L line), separates the one right-handed ($\pquant{S3}<0$) C point at $K'$ from the left-handed ($\pquant{S3}>0$) C points, one at $K$ and four near to $\Gamma$. In addition to the handedness, each C point is attributed with an index $I\in\frac{1}{2}\mathset{Z}$ which is determined by the winding of $\pquant{polar_azimuth}$ around this singularity. The numbers near to each C point indicate the value of $I\cdot\mathrm{sgn}(\pquant{S3})$. Their negative sum $-\sum_{c\in\mathcal{C}}I_c\cdot\mathrm{sgn}(\pquant{S3}(\vec{k}_c))$ yields the Chern number, in this example $C_n=-1$.
		(c) shows all the four combinations with one topologically stable C point, \abbr{ie} $I=\pm\frac{1}{2}$: $\pquant{polar_azimuth}$ winds up (down) for the upper left (right) and lower right (left) example, and the two upper (lower) C points are left-handed (right-handed). The resultant value for $I\cdot\mathrm{sgn}(\pquant{S3})$ is given there. These schematic plots are equivalent to \cref{fig:l_lines+chern}\,c.
	}
	\label{fig:c_points+chern}
\end{figure}

Next, we show that the Chern numbers can also be related to the properties of the so-called C points \cite{nye1983clines,dennis2009review_polarization_sing}. The C points are the points where the polarization gets perfectly circular. In other words, they correspond to the poles of the \name{Poincare} sphere, and they are the nodes of the scalar field \cite{konukhov_melnikov2001optical_vortices}
\begin{equation}
	\sigma = \pquant{S1} + \mconst{i} \pquant{S2} = \pquant{S0} \cos(2\pquant{polar_altitude})\,\mconst{e}^{2\mconst{i}\pquant{polar_azimuth}}
	\label{eq:definition_sigma_field}
\end{equation}
whose complex phase is directly related to the polarization azimuth $\pquant{polar_azimuth}$.

It turns out that the C points can be classified according to four different criteria \cite{berry_hannay1977umbilic_points,nye1983clines,dennis2008monstardom}. Below, we show that two of these criteria are relevant in determining the Chern number. The first relevant criterium is based on their handedness, \abbr{ie} whether the polarization is purely left-handed ($\pquant{S3}>0$) or right-handed ($\pquant{S3}<0$). The other relevant criterium is based on the so-called $I$ classification: the polarization azimuth $\pquant{polar_azimuth}$ is not well-defined at a C point, and the index $I$ counts its winding around the corresponding singularity (in counter-clockwise direction):
\begin{equation}
	I = \frac{1}{2\mconst{pi}} \oint \total{\pquant{polar_azimuth}} \in \frac{1}{2}\mathset{Z}
	\label{eq:definition_c_point_index}
\end{equation}
Its half-integer nature emerges from the fact that a half rotation of $\pquant{polar_azimuth}$ corresponds to a full rotation around the $\pquant{S3}$ axis, \abbr{ie} it is already sufficient to restore the original elliptical motion pattern. We note that a C point, by definition, emerges at any crossing of the contours $\pquant{S1}=0$ and $\pquant{S2}=0$. Whenever any random perturbation is introduced, all the crossings with $I\neq\pm\frac{1}{2}$ are splitted into several C points (the sum over all C points of the $I$ indexes is conserved in this process). In contrast, those C points with $I=\pm\frac{1}{2}$ are structurally stable, \abbr{ie} they will not split, and they cannot be created or destroyed spontaneously; pairwise creation and annihilation is however possible.

It turns out that also the sum of the $I$ indexes weighted by the corresponding handedness (the sign of $\pquant{S3}(\vec{k})$) is conserved for all these continuous transformations. In other words, such a weighted sum over all C points is a topological invariant. In the present setting where the polarization field is associated to the $n$-th energy band of a translationally invariant system, the topological invariant discussed above can be identified with the Chern number,
\begin{equation}
	C_n =
	- \sum_{c\in\mathcal{C}} I_c \cdot \mathrm{sgn}(S_3(\vec{k}_c)) =
	\sum_{c\in\mathcal{C}_-} 2I_c =
	- \sum_{c\in\mathcal{C}_+} 2I_c
	\label{eq:relation_chern+c_points}
\end{equation}
where $\mathcal{C}_+$ ($\mathcal{C}_-$) contains all the C points with $\mathrm{sgn}(S_3(\vec{k}_c))>0$ ($<0$) and $\mathcal{C}=\mathcal{C}_+\cup\mathcal{C}_-$.
% This relation is valid for all types of C points, in particular also topologically unstable C points with an index different from $\pm\frac{1}{2}$.

\Cref{eq:relation_chern+c_points} can be obtained from \cref{eq:relation_chern+l_lines} by using that the winding of the polarization azimuth equals the sum of the enclosed C point indices. The two last conversions in \cref{eq:relation_chern+c_points} are possible because the winding number along the boundaries of the Brillouin zone has to vanish (contributions on opposite edges will always cancel), so the unweighted index sum always vanishes: $\sum_{c\in\mathcal{C}}I_c=0$.

\begin{figure}[t!]
	\centering
	\includegraphics[scale=0.75]{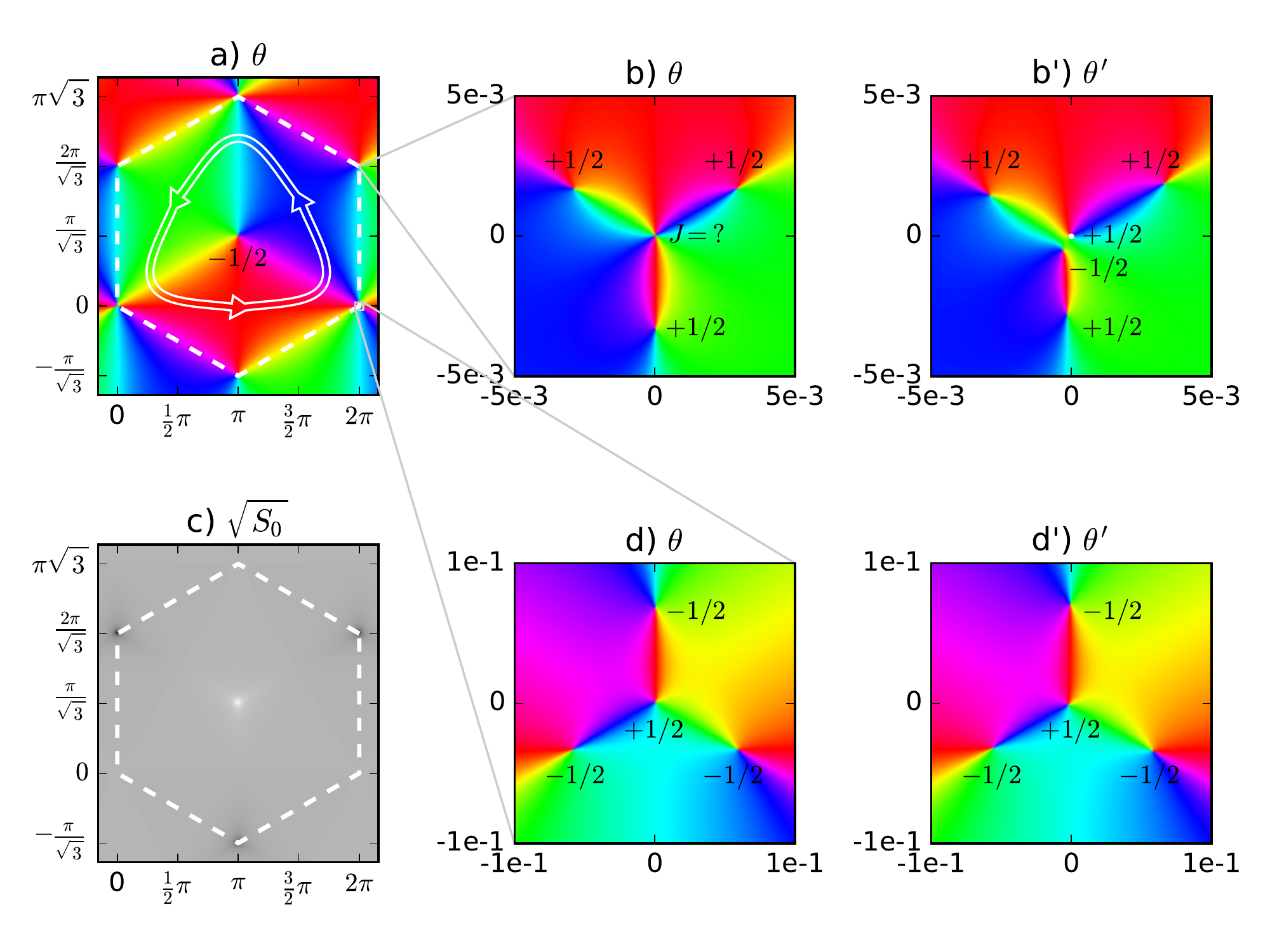}
	\caption{Interpretation of the Chern number in the presence of amplitude vortices.
		(a) is similar to \cref{fig:c_points+chern}\,a, but now for the field $\vec{\psi}^{(2)}(\vec{k})$ associated to the second band. (b,d) Zoom-ins at the $K$ and $\Gamma$ point, respectively. (b',d') Zoom-ins for a slightly modified Hamiltonian, see below. (c) Amplitude of $\vec{\psi}^{(2)}(\vec{k})$.
		The crucial difference with respect to \cref{fig:c_points+chern}\,a is the presence of an amplitude vortex, located directly at the $K$ point (\abbr{cmp} c). This point can be taken into account by an additional contribution $J$ to the Chern number (\abbr{cmp} \cref{eq:relation_chern+c_points+ampl_vortices}), but it is impossible to determine its value if no additional component of the Bloch state $\ket{\Phi_n(\vec{k})}$ is known.
		Alternatively, the value of $J$ can be understood intuitively in the following way: the amplitude vortex is unstable and under a small perturbation it decays into C points. Since the amplitude vortex in (b) is protected by rotational symmetry, it can be destroyed only by some modification to the Hamiltonian which breaks this symmetry; (b') and (d') show the resulting $\pquant{polar_azimuth}$ plots. Whereas nothing significant happens from (d) to (d'), subplot (b') shows how the amplitude vortex in (b) decays into two C points with contributions $-1/2$ and $+1/2$ (the C point which stays at the $K$ point is surrounded by a tiny L line, and hence left-handed). From this, it can be concluded that $J=0$. Note that $\pquant{S0}\cdot\mconst{e}^{2\mconst{i}\pquant{polar_azimuth}}=\sigma/\cos(2\pquant{polar_altitude})$ is discontinuous at C points, so $\pquant{S0}$ and $2\pquant{polar_azimuth}$ cannot simply be interpreted as the amplitude and the phase of a scalar field.
	}
	\label{fig:c_points+ampl_vortices+chern}
\end{figure}

Like \cref{eq:relation_chern+l_lines}, \cref{eq:relation_chern+c_points} applies only in the absence of amplitude vortices. Again, as we already discussed above in the context of L lines, such amplitude vortices are not topologically stable and do not appear generically. They can typically be eliminated by choosing a different projection. However, if needed, it is easy to extend \cref{eq:relation_chern+c_points} by their contributions:
\begin{equation}
	C_n = - \sum_{c\in\mathcal{C}} I_c \cdot \mathrm{sgn}(S_3(\vec{k}_c)) - \sum_{a\in\mathcal{A}} J_a
	\label{eq:relation_chern+c_points+ampl_vortices}
\end{equation}
where $\mathcal{A}$ consists of the labels for the amplitude vortices. The concrete value of $J_a$ cannot be deduced from the values of $\Psi_{s_1}^{(n)}(\vec{k})$ and $\Psi_{s_2}^{(n)}(\vec{k})$ alone, but it can be determined if an additional projection $\Psi_{s_3}^{(n)}(\vec{k})$ is known (which must not vanish at the same point $\vec{k}$). Alternatively, $J_a$ can be interpreted as the contributions of the C points into which the amplitude vortex decays under a small perturbation of the Hamiltonian. The treatment of amplitude vortices is illustrated in \cref{fig:c_points+ampl_vortices+chern}.

\section{Connection to other Methods and Proof}
\label{sec:connection+proof}

We will now present a proof for our recipe in \cref{sec:l_lines} to obtain the Chern number from the L lines. In addition, we will relate the schemes presented in \cref{sec:l_lines} and \cref{sec:c_points} with another, well-known technique to visualize the Chern number, the Skyrmion configuration method, and discuss common features and differences.

\subsection{Kohmoto method}

We will prove our main result formula, \cref{eq:relation_chern+l_lines}, for the Chern number by making a connection to the well-known analytical method for calculating the Chern numbers which is due to Khomoto \cite{kohmoto1985topological_invariant}. In Khomoto's method, one fixes a gauge by requiring that the overlap between the Bloch eigenstates of a particular band with a fixed state, for example $\ket{j,s}$, is a real positive number. When this prescription is well defined in the whole Brillouin zone, the Chern number can be interpreted as the flux of a curl piercing a closed surface (the Brillouin zone which is a torus), \abbr{cf} \cref{eq:definition_chern_number}. In this case, Stokes theorem ensures that the Chern number will be zero. Thus, in all topologically non-trivial cases the prescription will be ill-defined for one or more values of the quasi-momentum $\vec{k}$. These points are commonly referred to as obstructions. The reason why the gauge prescription is ill-defined at an obstruction is simply that the corresponding overlap matrix element vanishes there. In this situation, it is neccessary to divide the Brillouin zone into several regions where different gauge conditions are applied. Then, the Chern number is encoded in the phase mismatches $\varphi_{AB}$ at the interface between the different regions, $\exp(\mconst{i}\varphi_{AB}(\vec{k})):=\braket{\Phi_A(\vec{k})}{\Phi_B(\vec{k})}$ where $\ket{\Phi_{A,B}(\vec{k})}$ is the wavefunction in the gauge chosen in the regions $A$ and $B$, respectively. If the subdivision into regions is chosen such that there are no trijunctions, \abbr{ie} all the interfaces are closed loops, the Chern number is
\begin{equation}
	C_n = \frac{1}{2\mconst{pi}} \sum_{A\sqsubset B} \oint \total{\varphi_{AB}}
	\label{eq:kohmoto_method}
\end{equation}
where the sum is taken over interfaces between neighboring regions $A$, $B$ (where $A$ is enclosed by $B$), and the integral is along the respective interface line. In practice, an effective route to analytically calculate the Chern number is to choose one gauge $\ket{\Phi_B(\vec{k})}$, determine the corresponding obstructions, and, for each obstruction $j$, fix a suitable gauge $\ket{\Phi_{A_j}(\vec{k})}$ in a small region around it. This procedure reduces the task of calculating the Chern number to computing the phase mismatch in a few infinitesimally small regions. This often paves the way to an analytical treatment.

Our scheme discussed in \cref{sec:l_lines} can be viewed as a variation of Khomoto's method where rather than trying to fix the same gauge in the whole Brillouin zone and inserting patches where this does not work, we focus from the very beginning on two different gauge choices. For this purpose, we define the left- and right-handed components of an eigenstate $\ket{\Phi}$ (here, we omit the band index $n$) using the corresponding auxiliary polarization field from \cref{eq:definition_polarized_field}:
\begin{align}
	\pquant{psi_L}[\Phi] &:= \frac{1}{\sqrt{2}}(\Psi_{s_1}[\Phi]-\mconst{i}\Psi_{s_2}[\Phi]) &
	\pquant{psi_R}[\Phi] &:= \frac{1}{\sqrt{2}}(\Psi_{s_1}[\Phi]+\mconst{i}\Psi_{s_2}[\Phi])
	\label{eq:definition_psi_lr}
\end{align}
Demanding either $\pquant{psi_L}$ or $\pquant{psi_R}$ to be real, we obtain two distinct gauge conditions for the state $\ket{\Phi}$:
\begin{align}
	\pquant{psi_L}[\pquant{Phi_L}(\vec{k})] & \stackrel{!}{\in} \mathset{R+} &
	\pquant{psi_R}[\pquant{Phi_R}(\vec{k})] & \stackrel{!}{\in} \mathset{R+}
	\label{eq:left+right-handed_gauge}
\end{align}
We will fix the gauge using the first condition when the state $\ket{\Phi}$ is in the northern hemisphere of the \name{Poincare} sphere (corresponding to $\abs{\pquant{psi_L}}>\abs{\pquant{psi_R}}$, \abbr{ie} left-handed polarization), and the second one otherwise. This construction is possible because we have assumed that there are no amplitude vortices. We now show that the phase mismatch between these two gauges, defined by 
$\exp(\mconst{i}\varphi_\mathrm{RL}(\vec{k})):=\braket{\pquant{Phi_R}(\vec{k})}{\pquant{Phi_L}(\vec{k})}$, is directly related to the polarization azimuth:
$\varphi_\mathrm{RL}=2\pquant{polar_azimuth}$. For this purpose, we first apply $\braket{\Phi_\mathrm{R}}{\Phi_\mathrm{L}}=\conj{\pquant{psi_L}}\pquant{psi_R}/\abs{\conj{\pquant{psi_L}}\pquant{psi_R}}$.
% Note that the right-hand side is still gauge-invariant, \abbr{ie} independent of whether $\ket{\Phi}=\ket{\pquant{Phi_L}}$ or $\ket{\Phi}=\ket{\pquant{Phi_R}}$.
By inserting \cref{eq:definition_psi_lr} into the numerator and comparing the result with the definition of $\sigma$ in \cref{eq:definition_sigma_field} rewritten in terms of $\Psi_{s_1}$ and $\Psi_{s_2}$ using \cref{eq:definition_stokes_parameters}, we find that $\sigma=2\conj{\pquant{psi_L}}\pquant{psi_R}$. So, $\exp(\mconst{i}\varphi_\mathrm{RL}(\vec{k}))=\sigma/\abs{\sigma}=\exp(2\mconst{i}\pquant{polar_azimuth})$ according to the last equality in \cref{eq:definition_sigma_field}.

\subsection{Skyrmion number}

It has been discussed before \cite{fradkin2013field_theory_cond_mat,bernevig2013topo_insulators} that in a two-band model, the Chern number can be related to the skyrmion number: every eigenstate is represented by a Bloch vector $\vec{d}$ which is analoguous to the vector $(\pquant{S1},\pquant{S2},\pquant{S3})$ here; the skyrmion number
\begin{equation}
	\frac{1}{4\mconst{pi}} \iint_\mathrm{BZ} \vec{d}\cdot\bigg(\frac{\partial\vec{d}}{\partial x}\times\frac{\partial\vec{d}}{\partial y}\bigg)\,\total[^2]{k}
\end{equation}
counts how many times the mapping $\vec{d}(\vec{k})$ wraps the Bloch sphere, and is identical to the Chern number of the corresponding energy band. The skyrmion number has been developed in the context of magnetic skyrmions, quasiparticles which appear in certain magnetic materials\cite{nagaosa2013review_magnetic_skyrmions}.

In that sense, our method generalizes this ``skyrmion method'' to models with an arbitrary number of bands. In addition, it connects the Chern number to the theory of L lines and C points, and it provides an interpretation of the Chern number in terms of directly visible features.

\section{Application to Mechanics and Optics}
\label{sec:application}

\begin{figure}[b!]
	\centering
	\includegraphics{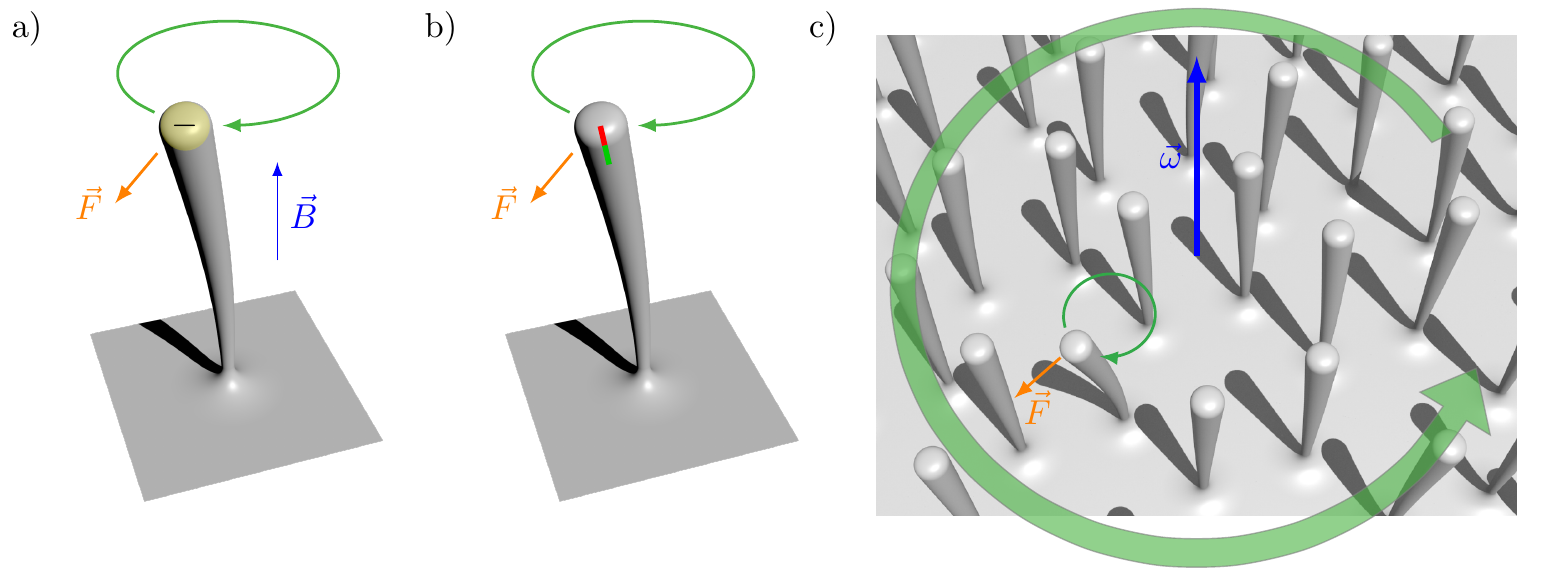}
	\caption{Several options to implement Zeeman splitting in a nanopillar array.
		a) Interaction between a charged nanopillar and a homogeneous external magnetic field $\vec{B}$ perpendicular to the surface.
		b) Precession induced by the angular momentum of a magnetic tip (\abbr{cmp} Einstein-de Haas effect\cite{einstein1915edh_theory,einstein_de_haas1915edh_experimental}) whose magnetic moment corresponds to the one of the magnet indicated in the figure.
		c) Global rotation of the nanopillar array. As observed in the co-rotating frame, there is the centrifugal force pointing directly away from the rotation axis (blue). Additionally, a relative movement of a pillar (indicated for one example) will induce a force perpendicular to its direction of motion, the Coriolis force, giving rise to Zeeman splitting.
	}
	\label{fig:nanopillars}
\end{figure}

The illustrative examples in \cref{fig:l_lines+chern,fig:c_points+chern,fig:c_points+ampl_vortices+chern} have actually been calculated for a specific model which has already been implemented for gyroscopic pendula \cite{nash2015gyroscopic_metamaterials} and proposed for photon-exciton polaritons\cite{jacqmin2014honeycomb_polaritons,sala2015soc_polaritons,nalitov2015soc_photonic_graphene,nalitov2015polariton_aqhe}. This tight-binding  model features a $\mathset{C2}$ polarization degree of freedom on a honeycomb lattice. The on-site polarization degree of freedom represents two in-plane directions of motion of a two-dimensional mechanical (\abbr{eg} gyroscopic) oscillator or two directions of the field distributions in the case of photon-exciton polaritons. Because of the two sublattices in a honeycomb lattice, there are in total four energy/frequency bands in such a tight-binding model. For a pair of sites, the longitudinal (transversal) mode is defined such that it is parallel (perpendicular) to the connecting line. The longitudinal (transversal) modes of nearest neighbor sites are coupled with coupling strength $t_\mathrm{L}$ ($t_\mathrm{T}$). The toy model for this interaction in mechanics are two-dimensional mass-spring networks. In this setting, additional on-site Zeeman splitting -- breaking the time-reversal symmetry -- gives rise to AQHE edge channels\cite{kariyado2015mech_graphene,nalitov2015polariton_aqhe}.

Because of its simple conception and the possibility to directly observe the polarization patterns in a mechanical system, such a system is a promising platform to demonstrate the determination and visualization of the Chern number with the technique described in this article. Whereas a macroscopic implementation of this model seems straightforward, a mesoscopic or microscopic one is more challenging due to the difficulties to fabricate springs at these lengthscales. However, there might be some more easily fabricated systems which intrinsically have this kind of interaction, \abbr{eg} arrays of nanopillars \cite{paulitschke2011phd,paulitschke2013gaas_nanopillars}. Unpublished experimental data \cite{weig_priv_comm} indicate that the motion of neighboring nanopillars is coupled, and the described coupling type is the only linear interaction model which respects all the present symmetries.

Conceptually, there are different options to implement Zeeman splitting including charged nanopillars in a homogeneous magnetic field, the conserved angular momentum \cite{nash2015gyroscopic_metamaterials} of magnetic tips in analogy to the Einstein-de Haas effect\cite{einstein1915edh_theory,einstein_de_haas1915edh_experimental}, and the Coriolis force \cite{kariyado2015mech_graphene} (\abbr{cmp} \cref{fig:nanopillars}). However, order-of-magnitude estimates indicate that only the latter one can be strong enough to overcome the effect of dissipation and disorder.

Another possible application, in the context of optics, are dielectric materials with a discrete translational symmetry in two dimensions; chiral edge channels analoguous to the QHE have already gained experimental verification in photonic crystals \cite{wang2009topo_elm_states}. In these systems, the (complex) electric field $\vec{E}(x,y)$ takes over the role of the electron wave function. If arbitrary quasimomentum states can be excited and the resultant polarization of the electric field $\vec{E}(x_0,y_0)$ can be measured at one single point $(x_0,y_0)$ which is fixed previously, mapping these polarization patterns onto the Brillouin zone allows to directly read off the Chern number. Note that although the degree of polarization of the field $\vec{E}$ is only two, the dimension of the Bloch space is infinite due to the dependency on the two (continuous) coordinates $x$ and $y$. This is one example where the generalization to an arbitrary number of bands, compared to the Skyrmion method (see \cref{sec:connection+proof}), becomes important.

\section{Measuring the Chern number}

Due to its outstanding role in the description of topologically protected edge state, the measurement of the Chern number may be a crucial point in an experimental analysis of a Chern insulator.

One possible approach to obtain the Chern numbers is to measure the quantized Hall conductance in the band gaps. For example, this measurement works with very high precision in the original setup, the quantum Hall effect \cite{klitzing1980qhe,klitzing1986review_qhe}. However, there might be some situations in which wrong results are obtained due to technical limitations, \abbr{eg} if edge states in one band gap exist in principle, but they have a too large penetration length compared to the system size to be detectable. Or, if there is only a local, but not a global band gap, the Chern number is well-defined, but cannot be measured this way. More importantly, this method is by a fundamental reason restricted to fermionic systems as it relies on the quantized Hall conductance. Since there is no equivalent quantity in a bosonic system, different approaches to measure the Chern number for those had to be developed \cite{aidelsburger2015measure_chern_ultracold_atoms,price2016measure_chern_center_of_mass,mittal2016measure_chern_photonic_system,bardyn2014measure_chern_photonic_lattices,schroer2014measure_topo_transition}.

Our method could serve as an alternative to these techniques as it offers a very direct approach: experimental techniques like Fourier transform spectroscopy or the excitation with quasimomentum modes give full access to the Bloch states; this is sufficient input to identify the L lines and C points, and to determine their relevant properties. Therefore, straightforward application of the schemes in \cref{sec:l_lines,sec:c_points} can also be used to measure the Chern numbers for any type of Chern insulator.

Its universality and directness could be a great advantage in comparison to other strategies. We also emphasize that, as opposed to, for example, a numerical calculation using a band structure simulation (there are both analytical \cite[\abbr{sec} III]{kohmoto1985topological_invariant} and numerical \cite{fukui2005chern_number_efficient_method} methods), our method requires neither precise knowledge of the actual system parameters nor even the understanding of the underlying microscopic mechanisms.

\section{Conclusion}
\label{sec:conclusion}

The central aspect of this article is the connection between two topological concepts in modern physics: we have related the Chern number, one of the central quantities in the context of the QHE and the AQHE, to L lines and C points, the structurally stable objects known from random fields in polarized optics. What we obtain is a graphical interpretation of the Chern number which makes this abstract quantity more tangible.

The term polarization is fundamental in the geometric description of oscillating electromagnetic fields. Even more, we can directly see what it means to a mechanical system by looking at the trajectories. This intuition makes these two classes of systems play an exclusive role for the scheme presented here.

Besides the visualization, we have also discussed the possibility to use this technique as a tool to measure the Chern number in an experiment.

\section*{References}

\bibliography{references}

% \section*{Supplementary}
% 
% 
% ?:
% \begin{itemize}
% 	\item full topological phase diagram
% 	\item numbers for nanopillars
% 	\item derivations L line to C point, amplitude vortices
% 	\item Chern number from azimuth and altitude
% \end{itemize}

\end{document}